\documentclass[twocolumn]{aastex631}

\turnoffeditone
\shorttitle{Polarization Spectrum of NIR ZL Observed with CIBER}
\shortauthors{Takimoto et al.}

\graphicspath{{./}{figures/}}

\begin{document}

\title{POLARIZATION SPECTRUM OF NEAR INFRARED ZODIACAL LIGHT \\
OBSERVED WITH CIBER}

\author{Kohji Takimoto}
\affiliation{Department of Physics, School of Science and Technology, Kwansei Gakuin University, Hyogo 669- 1337, Japan}
\author{Toshiaki Arai}
\affiliation{Department of Infrared Astrophysics, Institute of Space and Astronautical Science (ISAS), Japan Aerospace Exploration Agency (JAXA), Sagamihara, Kanagawa 252-5210, Japan}
\author{Shuji Matsuura}
\affiliation{Department of Physics, School of Science and Technology, Kwansei Gakuin University, Hyogo 669- 1337, Japan}
\author{James J. Bock}
\affiliation{Department of Physics, California Institute of Technology, Pasadena, CA 91125, USA}
\affiliation{Jet Propulsion Laboratory (JPL), National Aeronautics and Space Administration (NASA), Pasadena, CA 91109, USA}
\author{Asantha Cooray}
\affiliation{Center for Cosmology, University of California, Irvine, Irvine, CA 92697, USA}
\author{Richard M. Feder}
\affiliation{Department of Physics, California Institute of Technology, Pasadena, CA 91125, USA}
\author{Phillip M. Korngut}
\affiliation{Department of Physics, California Institute of Technology, Pasadena, CA 91125, USA}
\author{Alicia Lanz}
\affiliation{The Observatories of the Carnegie Institution for Science, Pasadena, CA 91101, USA}
\author{Dae Hee Lee}
\affiliation{Korea Astronomy and Space Science Institute (KASI), Daejeon 305-348, Korea}
\author{Toshio Matsumoto}
\affiliation{Department of Infrared Astrophysics, Institute of Space and Astronautical Science (ISAS), Japan Aerospace Exploration Agency (JAXA), Sagamihara, Kanagawa 252-5210, Japan}
\author{Chi H. Nguyen}
\affiliation{Department of Physics, California Institute of Technology, Pasadena, CA 91125, USA}
\author{Yosuke Onishi}
\affiliation{Department of Infrared Astrophysics, Institute of Space and Astronautical Science (ISAS), Japan Aerospace Exploration Agency (JAXA), Sagamihara, Kanagawa 252-5210, Japan}
\author{Kei Sano}
\affiliation{Department of Space Systems Engineering, School of Engineering, Kyushu Institute of Technology, Fukuoka 804-8550, Japan}
\author{Mai Shirahata}
\affiliation{Department of Infrared Astrophysics, Institute of Space and Astronautical Science (ISAS), Japan Aerospace Exploration Agency (JAXA), Sagamihara, Kanagawa 252-5210, Japan}
\author{Aoi Takahashi}
\affiliation{Astrobiology Center, National Institutes of Natural Sciences, Mitaka, Tokyo 181-8588, Japan}
\author{Kohji Tsumura}
\affiliation{Department of Natural Science, Faculty of Science and Engineering, Tokyo City University, Setagaya, Tokyo, 158-8557, Japan}
\author{Michael Zemcov}
\affiliation{Center for Detectors, School of Physics and Astronomy, Rochester Institute of Technology, Rochester, NY 14623, USA}
%% Note that the \and command from previous versions of AASTeX is now
%% depreciated in this version as it is no longer necessary. AASTeX 
%% automatically takes care of all commas and "and"s between authors names.

%% AASTeX 6.31 has the new \collaboration and \nocollaboration commands to
%% provide the collaboration status of a group of authors. These commands 
%% can be used either before or after the list of corresponding authors. The
%% argument for \collaboration is the collaboration identifier. Authors are
%% encouraged to surround collaboration identifiers with ()s. The 
%% \nocollaboration command takes no argument and exists to indicate that
%% the nearby authors are not part of surrounding collaborations.

%% Mark off the abstract in the ``abstract'' environment. 
\begin{abstract}

We report the first measurement of the zodiacal light (ZL) polarization spectrum in the near-infrared between 0.8 and 1.8 $\mu$m.
Using the low-resolution spectrometer (LRS) on board the Cosmic Infrared Background Experiment (CIBER), calibrated for absolute spectrophotometry and spectropolarimetry, we acquire long-slit polarization spectral images of the total diffuse sky brightness towards five fields.
To extract the ZL spectrum, we subtract contribution of other diffuse radiation, such as the diffuse galactic light (DGL), the integrated star light (ISL), and the extragalactic background light (EBL).
The measured ZL polarization spectrum shows little wavelength dependence in the near-infrared and the degree of polarization clearly varies as a function of the ecliptic coordinates and solar elongation. 
Among the observed fields, the North Ecliptic Pole shows the maximum degree of polarization of $\sim$ 20$\%$, which is consistent with an earlier observation from the Diffuse Infrared Background Experiment (DIRBE) aboard on the Cosmic Background Explorer (COBE).
The measured degree of polarization and its solar elongation dependence are reproduced by the empirical scattering model in the visible band and also by the Mie scattering model for large absorptive particles, while the Rayleigh scattering model is ruled out.
All of our results suggest that the interplanetary dust is dominated by large particles.

\end{abstract}


\section{Introduction} \label{sec:intro}

The zodiacal light (ZL) arises from sunlight scattered by the interplanetary dust  (IPD) in the optical and the near-infrared \edit1{($<3~\mu$m)}, and from thermal emission from IPD in the mid- and far-infrared \edit1{($>3~\mu$m)}.
Measuring the ZL is important for understanding the structure of the IPD distribution and physical properties of the IPD, such as their size distribution, composition, shape, and complex refractive index.
In addition, information obtained from ZL measurements is important for future studies of dust disks of exoplanetary systems. 
\edit1{Most of the ZL brightness is from the IPD, with typical dust grain sizes from 1 $\mu$m and 100 $\mu$m, probably even less than 1 $\mu$m \citep{1985Icar...62..244G,reach1988zodiacal}}.
Because the IPD grains of these sizes fall into the Sun \edit1{due to Poynting-Robertson drag \citep{burns1979radiation} within the age of the Solar System}, IPD dating to the early solar system is unlikely to exist today.
\edit1{\citet{wyatt1950poynting} found that all particles with a radius $\leq$ 1 cm, which were initially in a sphere with a radius of 1 A.U. centered on the Sun, fall into the Sun with a period of 28 million years.}
A continuous supply of IPD particles is necessary to maintain the zodiacal cloud.

The source of the IPD is assumed to comprise both comets and asteroids, but the relative importance of each has not been settled.
\edit1{Interstellar dust also contributes to the IPD at a level estimated to be less than 10\% \citep{srama2011cosmic,rowan2013improved}.}
\citet{2010ApJ...713..816N,Nesvorn__2011} compared their dynamical simulations of the IPD with the Infrared Astronomical Satellite (IRAS) data suggesting that 85 $\sim 95\%$ of the observed mid-infrared emission is 
produced by particles from Jupiter-family comets (JFCs) and $< 10\%$ by dust from long-period comets. 
\citet{2006AJ....132.1354F} showed that the geometric albedo \edit1{of the nucleus} of comet 162P/Siding Spring (P/2004 TU$_{12}$), which is largest radii among known JFCs, is $0.059 \pm 0.023$ in H band, $0.037 \pm 0.014$ in R band, and $0.034 \pm 0.013$ in V band. 
\citet{2002Sci...296.1087S} also observed \edit1{the nucleus of} the JFC 19P/Borrelly and derived a geometric albedo of $0.03 \pm 0.05$.
The IPD is also known to be a low albedo \citep{1998ApJ...508...44K} and has similarities to the comet nucleus.
\citet{Yang_2015} derived the spectral gradient of IPDs as $S'=8.5 \pm 1.0\% / 100$ nm at 460 nm, and combining with the albedo showed that $>90\%$ of the IPDs originate from comets or D-type asteroids.
The ZL spectra measured by the Near Infrared Spectrometer (NIRS) on board the Infrared Telescope in Space (IRTS) and the Cosmic Infrared Background Experiment (CIBER) show silicate-like features at 0.9 and 1.6 $\mu$m \citep{1996PASJ...48L..47M,2010ApJ...719..394T}, comparable to fresh and active comets or S-type asteroids.
A combination of amorphous and crystalline silicate grain features is found in ZL spectrum between 9 and 11$\mu$m \citep{1998EP&S...50..507O, 2003Icar..164..384R, 2009ASPC..418..395O}.
\edit1{Crystalline silicate features have been observed in comet C/1995 O1 Hale-Bopp \citep{wooden1999silicate,wooden2000mg}.}
\edit1{Therefore, the silicate feature of the ZL has not been the key to determine the fraction of cometary and asteroid dusts in the IPD.}

\edit1{Polarization of the ZL is another observable quantity to explore the source of IPD through the dust composition.}
The ZL scattered by an optically thin cloud of IPDs presents a systematic polarization. 
The linear polarization degree $P$ of sunlight scattered by IPD surfaces is usually defined as the difference between the intensities polarized along the planes perpendicular, $I_{\perp}$, and parallel, $I_{\parallel}$, to the scattering plane:
$$P=\frac{I_{\perp}-I_{\parallel}}{I_{\perp}+I_{\parallel}}.$$
According to \citet{1998ApJ...508...44K}, the ZL intensity $I_{\lambda}$ observed at wavelength $\lambda$ as the integral along the line of sight $s$ is expressed as:
$$
	 I_{\perp, \lambda} = \int{n(X, Y, Z)A_{\lambda}F^{\odot}_{\lambda}\Phi_{\perp, \lambda}(\Theta)}ds,$$
$$
	 I_{\parallel, \lambda} = \int{n(X, Y, Z)A_{\lambda}F^{\odot}_{\lambda}\Phi_{\parallel, \lambda}(\Theta)}ds,
$$
where $n(X, Y, Z)$ is the three-dimensional density of the IPD, $A_{\lambda}$ is the albedo at wavelength $\lambda$, $F^{\odot}_{\lambda}$ is the solar flux, and $\Phi_{\lambda}(\Theta)$ is the phase function at scattering angle $\Theta$.
Since the polarization properties of scattered light depend on the scattering angle and the composition of the IPD, the origin of the IPD can be investigated by the polarization spectrum of ZL at various fields.
For example, the cometary dust ejected from comets is studied by polarization measurements. 
\citet{zubko2014dust} studied dust in comet C/1975 V1 (West) and reproduced the polarization measurement by models with Mg-rich silicate and amorphous carbon. 
\citet{2009Icar..199..129L} also studied dust of comet C/1995 O1 Hale-Bopp and 1P/Halley based on the polarization measurement and implied that the cometary dust can be explained by mixture of a non-absorbing silicate-type material and a more absorbing organic-type material.

A few studies report polarization measurements of the ZL.
\citet{1998A&AS..127....1L} summarized the ZL polarization measured from space in the visible and the near-infrared, and the degree of polarization is largely dependent on the helio-ecliptic longitude to first order.
At visible wavelengths, \edit1{integrated line of sight polarization on the sky varies from -3\% to +20\% depending on the elevation and azimuth of observation,} and shows little wavelength dependence \citep{1980IAUS...90...19W, 1979A&A....74...15P, 1982A&A...105..364L}.
In the near-infrared, only \citet{1994ApJ...431L..63B} has measured the polarization of the ZL from space by the Diffuse Infrared Background Experiment (DIRBE) aboard on the Cosmic Background Explorer (COBE) in discrete photometric bands at 1.25, 2.2, and 3.5 $\mu$m.
This result shows that the degree of polarization of the ZL is about 10 $\sim 20\%$ at solar elongation $\epsilon$ = 90$^\circ$\edit1{, and tends to decrease toward longer wavelengths}.

In this paper, we report a ZL polarization spectrum measurement from the Low Resolution Spectrometer (LRS) onboard CIBER. 
The purpose of polarization observation by LRS was to separate polarized ZL from \edit1{presumably} unpolarized EBL, and to identify spectral features by studying the wavelength dependence of polarization.
Our result is the first measurement of the ZL polarization spectrum in the near-infrared.

\section{LOW RESOLUTION SPECTROMETER/CIBER} 

\begin{figure*}[htbp]
\epsscale{0.7}
%\vskip -7cm
\plotone{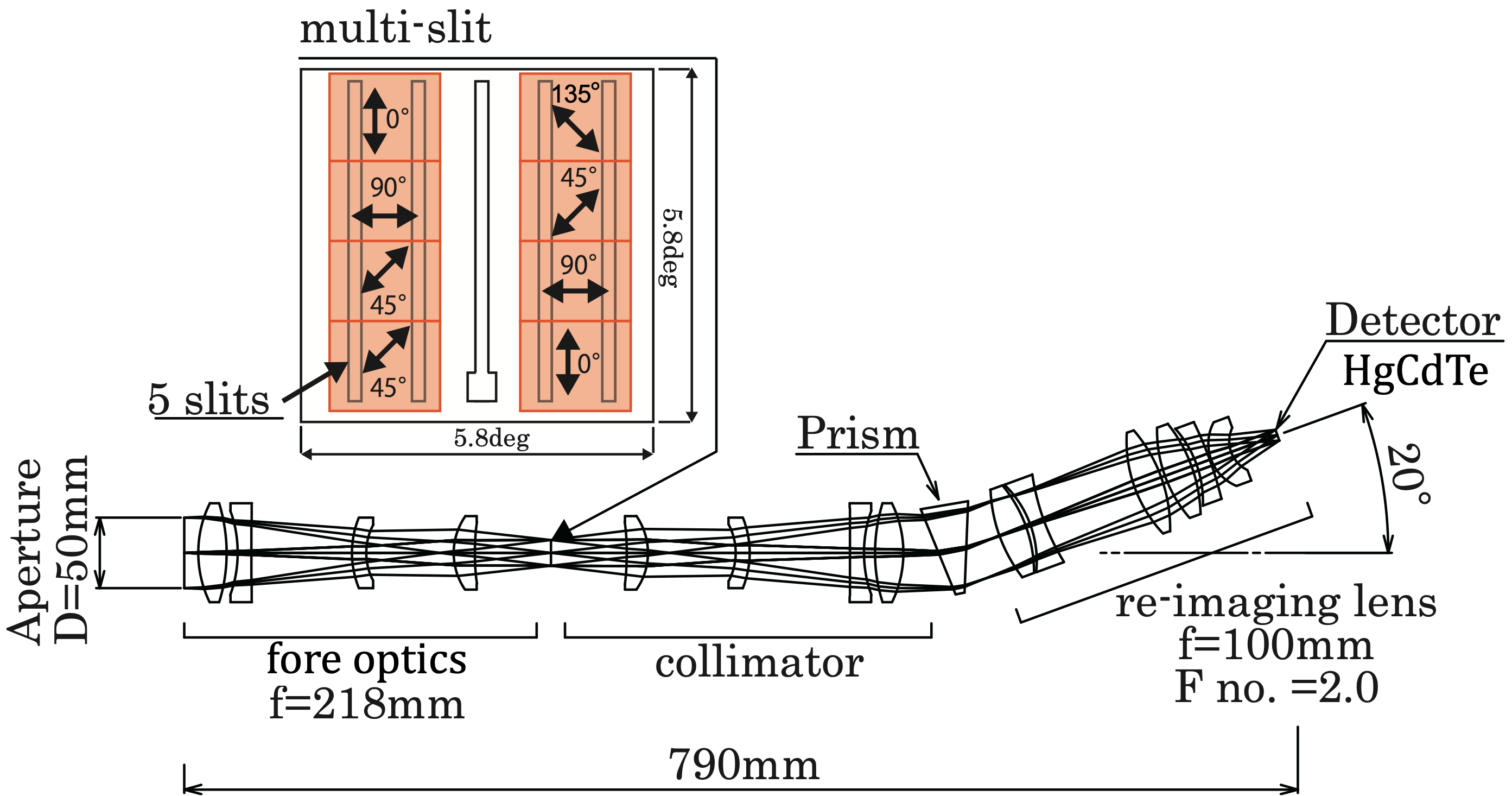}
%\vskip 7cm
\caption{The LRS optics and wire-grid polarizing films. 
The LRS has five slits.
\edit1{A notch at the bottom of the center slit is for the focus adjustment experiments in the laboratory.}
The wire-grid polarizing films (orange) are installed on the slit mask except for the center slit.
The transmittance axes are indicated by the black arrows.
The direction of the transmittance axes are defined as $\theta = 0^\circ, 45^\circ, 90^\circ, 135^\circ$.}
\label{fig:polarizer}
\end{figure*}

\begin{deluxetable*}{cccccccc}
\tabletypesize{\scriptsize}
% \rotate
\tablecaption{Observed Fields of the CIBER Third Flight}
\tablewidth{0pt}
\tablehead{\colhead{} & \colhead{Exposure} & \colhead{Payload's} & \colhead{Ecliptic} & \colhead{Galactic} & \colhead{Equatorial} &\colhead{Solar Elong.} & \colhead{ZL model} \\
\colhead{Field Name} & \colhead{Time} & \colhead{Altitude} & \colhead{($\lambda$, $\beta$)} & \colhead{($l$, $b$)} & \colhead{($\alpha$, $\delta$)} &\colhead{$\epsilon$} & \colhead{$\lambda I_{\rm ZL_{model}}$} \\
\colhead{} & \colhead{(s)} & \colhead{(km)} & \colhead{(deg)} & \colhead{(deg)} & \colhead{ (deg)} &\colhead{(degree)} & \colhead{(nW~m$^{-2}$~sr$^{-1}$)}}
\startdata
Lockman Hole & 47 & 202-265 & (135.42, 45.49) & (149.41, 51.97) & (161.43, 58.21) & 118.8  & 330.23\\
SWIRE/ELAIS-N1& 45 & 284-315 &(209.32, 72.32)&(84.31, 44.71)& (242.81, 54.59) & 105.7  & 269.12 \\
North Ecliptic Pole (NEP)  & 53 & 320-324 & (311.48, 89.62) & (96.06, 29.56) & (270.63, 66.28) & 90.0  & 280.55 \\
Elat30 ($\beta$ = 30 degree) & 26 & 319-306 & (234.38, 29.25) & (18.64, 44.89) & (236.98, 9.57) &  124.1 & 403.53 \\
BOOTES-B & 50 & 296-244 & (200.35, 44.82) & (55.13, 68.06) & (217.30, 33.27) & 132.3 & 328.47\\
\enddata
\tablenotetext{a}{The coordinate systems are based on J2000.}
\tablenotetext{b}{The ZL brightness is estimated by using the ZL model \citep{1998ApJ...508...44K} at 1.25 $\mu$m.}
\label{tbl:field3}
%% Text for table notes should follow after the \enddata but before
%% the \end{deluxetable}. Make sure there is at least one \tablenotemark
%% in the table for each \tablenotetext.
\end{deluxetable*}
CIBER, designed to study the diffuse near-infrared emission above the Earth's atmosphere  \citep{2013ApJS..207...31Z}, housed three instruments  including a two broad-band imagers \citep{2013ApJS..207...32B}, a narrow band spectrometer \citep{2013ApJS..207...34K}, as well as the LRS designed to measure the spectrum of diffuse light in 0.8$\leq \lambda \leq$ 1.8~$\mu$m \citep{2013ApJS..207...33T} with a wavelength resolution of $\lambda / \Delta \lambda~=~15$-30.
Fore optics of the LRS brings an image of the sky to focus on a mask containing 5 slits as shown in Figure \ref{fig:polarizer}, 
each spanning a field of view (FOV) of 5$^\circ \times$2.7$'$ sampled by a 256$\times$256 pixel HgCdTe array.
This array hosts a cold shutter assembly for dark current measurement.

CIBER conducted observations four times, on February 2009, July 2010, March 2012, and June 2013.
The payload was successfully recovered and refurbished after the first three flights.
Because the polarizers were only installed in the third flight, we mainly use the third flight data in this paper.
At the third flight, we used NASA Black Brant IX\footnote{For details on the launch vehicles, Sounding rocket handbook (http://sites.wff.nasa.gov/code810/files/SRHB.pdf).} two-stage vechicles launched from the White Sands Missile Range in New Mexico, USA.
The apogee on the flight was 330km, providing a total exposure time of  $\sim$ 240 seconds.
The raw data, which are non-destructively sampled by the integrating detectors,
were telemetered to the ground from  the rocket during the flight.
The celestial attitude control system achieved a pointing stability of $<$ 8$^{\prime \prime}$.
Details about the CIBER payload and flight performance are written in \citet{2013ApJS..207...31Z}.
See \citet{2013ApJS..207...33T} and \citet{arai2015measurements} for details of the LRS.

To measure the polarization spectrum, we installed wire-grid polarization film \footnote{Manufactured by Asahikasei E-materials corporation.} on the slit mask.
The wire-grid polarization films with different transmittance axis were installed on eight different regions of the slit mask (Figure \ref{fig:polarizer}).
Although the FOV of each polarizer is different, the ZL is smoothly distributed over the entire FOV and small field-to-field fluctuations if any can be corrected for.
Therefore, if the integrated star light (ISL) and diffuse galactic light (DGL) can be removed, ZL polarization can be measured with this instrument.

The phase angles of the transmittance axis of the polarizer are labeled as $\theta$~=~$0^\circ, 45^\circ, 90^\circ, 135^\circ$ (Figure \ref{fig:polarizer}).
\edit1{The relative angles of the four LRS polarizers are determined with an accuracy of $\pm0.1^\circ$ and are taken into account when estimating the polarization bias.}
Note that the polarizing film in the lower left corner was mistakenly installed with the transmission axis rotated by 90 degrees during the final installation, but it was calibrated and launched as it was.
Because the polarizing films were not installed on the center slit, the total spectrum of the sky brightness was also measured.

The observed fields are listed in Table \ref{tbl:field3}.
For the ZL analysis, the fields are selected based on ecliptic coordinates and solar elongation.
The low ecliptic latitude field, Elat30, shows high ZL brightness.
On the other hand, the high ecliptic latitude fields exhibit low ZL brightness.
The North Ecliptic Pole (NEP) field, where the solar elongation is 90$^\circ$, is selected because the polarization is expected to be maximum.
As the solar elongation increases, the polarization is expected to decrease.

% \newpage
\section{INSTRUMENTAL CALIBRATION FOR POLARIMETRY}\label{seq:cal}
We describe the wavelength calibration, surface brightness calibration, and flat field correction, and polarization calibration in this section.
The wavelength calibration, surface brightness calibration, and flat field correction are similar to \citet{arai2015measurements}. 
\subsection{Wavelength Calibration}
We conduct the wavelength calibration, which measures the relationship between the incident monochromatic light and a position on the detector array.
For spectral calibration in the laboratory, we use two different light sources consisting of the SIRCUS (spectral irradiance and radiance responsivity calibrations using uniform sources) laser facility and a standard quartz-tungsten-halogen lamp coupled to a monochrometer. 
To illuminate the LRS, both light sources are coupled via fiber to an integrating sphere 20 cm in diameter.
After exposure to a monochromatic light source, the detected signal is fitted with a Gaussian function.
The accuracy of the wavelength calibration is calculated to be $\pm$1 nm by combining the center of this Gaussian function with the externally determined wavelength of the incident light.
\subsection{Surface Brightness Calibration}
The LRS sensitivity requirement is $\rm \sim 10~nW~m^{-2}~sr^{-1}$, corresponding to 4\% level of the sky brightness.
To find the conversion between the photocurrent and the surface brightness, we use two different light sources consisting of the SIRCUS laser facility and a super-continuum laser (SCL). Both light sources coupled to a 20 cm diameter integrating sphere through a fiber illuminating the LRS aperture.
\edit1{The setup for surface brightness calibration is basically the same as for polarization calibration shown in Figure \ref{fig:cal_setup}, but without the external polarizer (polarization holder).}
The calibration factor is calculated from the measurement results of both light sources.
The statistical uncertainty of 1 $\sigma$ is estimated to be less than 0.1\% from the variance over all pixels of the detector array. The measured calibration factor is consistent within a 3$\%$ r.m.s variation.
\subsection{Flat Field Correction}
In order to correct for the spatial fluctuations generated by the detector's nonuniform response, we use laboratory measurements to construct the flat field.
We use three different light sources consisting of the SIRCUS laser facility, SCL, and a standard quartz-tungsten-halogen lamp with a solar-like filter. 
All light sources are coupled via fiber to an integrating sphere with a diameter of 10 or 20 cm, which then illuminates the aperture of the LRS.
By combining various light sources and integrating spheres, we can quantify the systematic uncertainty of the flat field correction. 
The illumination patterns of the spheres are uniform with an accuracy of less than 1\%.
\subsection{Polarization Calibration}
\begin{figure}[bp]
\epsscale{1.2}
\plotone{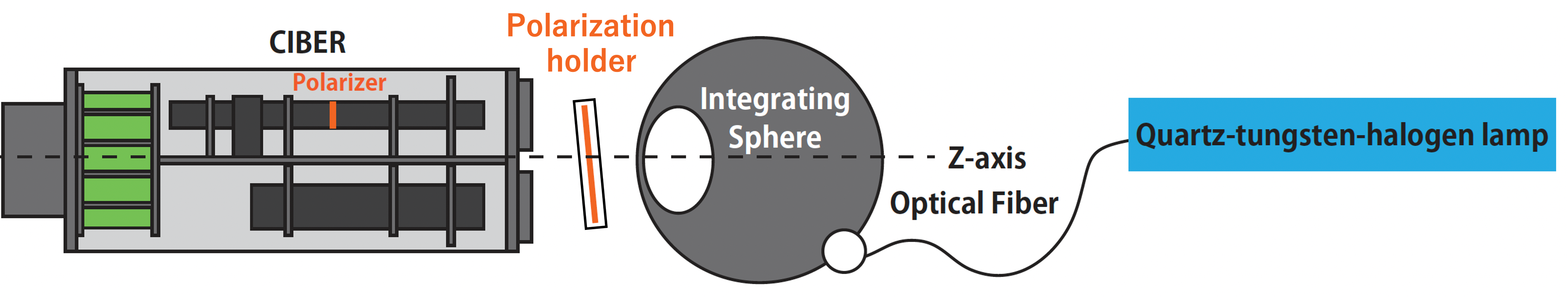}
\caption{Schematic view of the setup for the polarization calibration.
}
\label{fig:cal_setup}
\end{figure}
\begin{figure}[htbp]
\epsscale{1.2}
\plotone{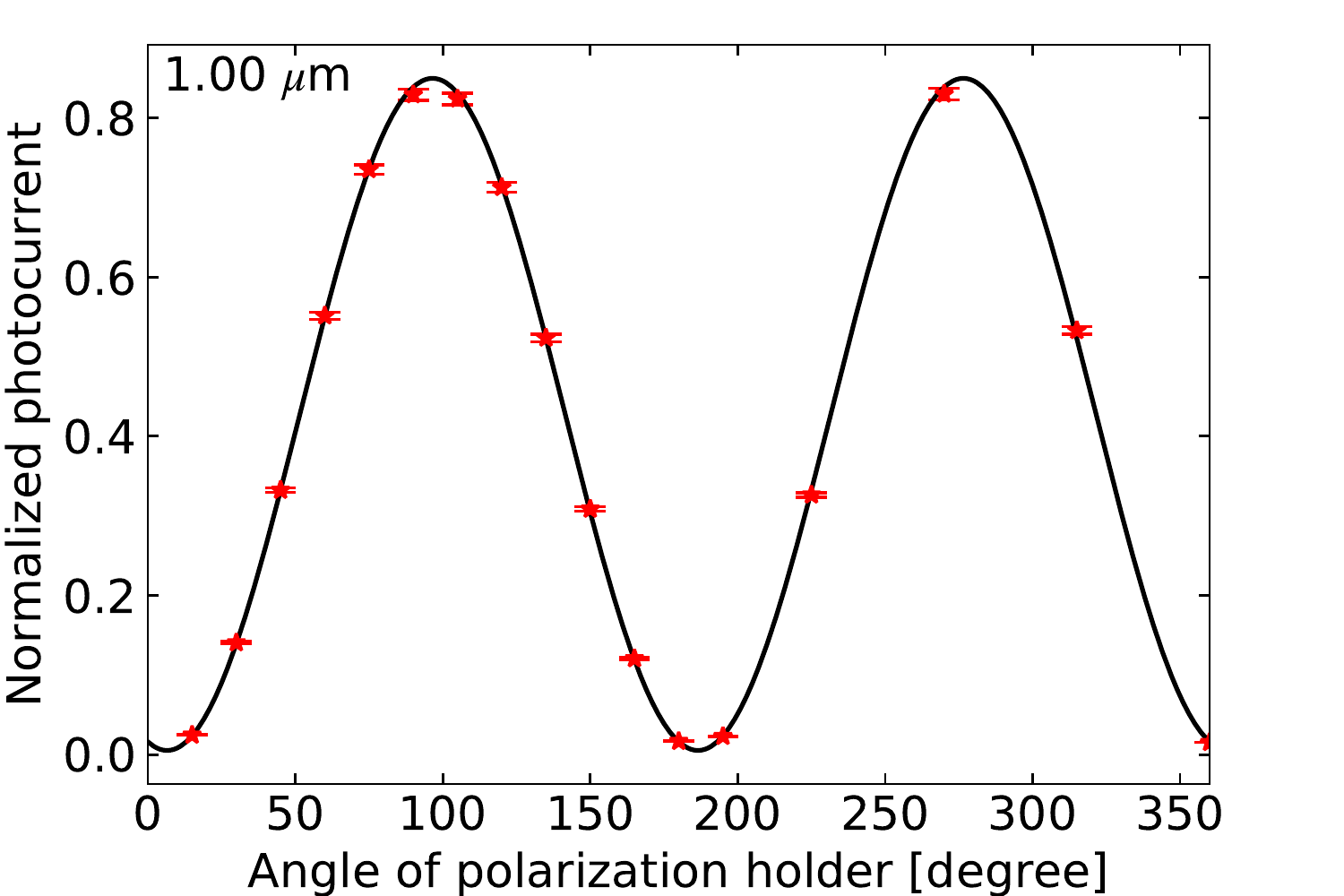}
\caption{Measured photocurrent through the polarization filter of $\theta = 0^\circ$ as a function of angle of polarization holder $\phi$ at $\lambda=$1.00 $\mu$m.
Because the power of the light source is not stable, the y-axis is normalized by the photocurrent of the center slit.
The red asterisks indicate measured data.
The black curve indicates best-fit by Equation \ref{eq:pol_def_lab} with the two variables $I_{{\rm pol}}$ and $I_{{\rm mean}}$.
}
\label{fig:sinfit_1um}
\end{figure}
\begin{figure}[tbp]
\epsscale{1.2}
\plotone{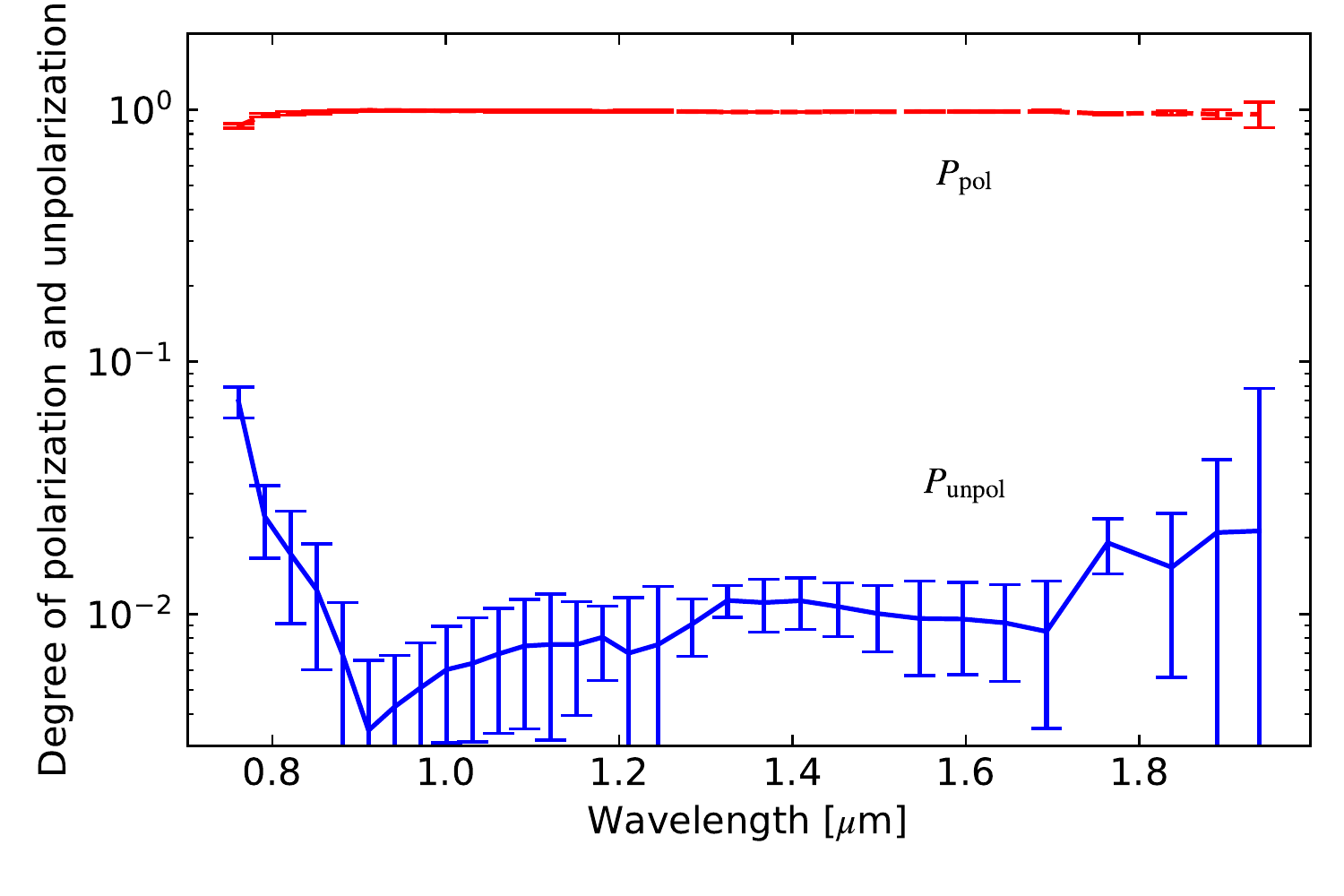}
\caption{The degree of polarization $P_{\rm pol}$ and the degree of unpolarization $P_{\rm unpol}$ of LRS measured in the laboratory.
The red dashed line indicates the degree of polarization defined by Equation \ref{eq:degree_of_polarization}.
The blue solid line indicates the degree of unpolarization.
Each error bar represents the total uncertainty due to the fitting of Equation \ref{eq:pol_def_lab} and polarization calibration.}
\label{fig:lab_test}
\end{figure}
In order to characterize the polarization measurement, we conduct the polarization calibration in the laboratory.
This polarization calibration determined the transmittance, the extinction ratio, and the relative angle of the LRS polarizer in the laboratory.
The extinction ratio quantifies the performance of the polarizer and is expressed as
\begin{equation}
r_{e} = \frac{P_{\rm pol}}{P_{\rm unpol}},
\label{eq:extinction_ratio}
\end{equation}
where $P_{\rm pol}$ is the polarization degree of the light passing through the polarizer, and $P_{\rm unpol}$ is the degree of the unpolarization $P_{\rm unpol}=1-P_{\rm pol}$.
The extinction ratio of the polarizer for parallel light was known as $>$ 1000 in advance, but not for convergent light.
The LRS polarizer operates with converging rays, so its extinction ratio needs to be measured in the laboratory before the flight.

A schematic view of the setup for the polarization calibration is shown in Figure \ref{fig:cal_setup}.
A standard quartz-tungsten-halogen lamp is used as light source and is coupled to an integrating sphere through a fiber illuminating the LRS aperture.
A 100 mm diameter wire-grid polarizing film is settled between the LRS aperture and the integrating sphere, which is installed on a rotational polarization holder.
The polarization holder is rotated around the optical axis of the LRS by 10$^\circ$ step. 

In Figure \ref{fig:sinfit_1um}, we show the normalized photocurrent measured with the LRS as a function of the angle of the polarization holder $\phi$\edit1{ at the wavelength band of 1.00 $\mu$m}. 
To quantify the degree of polarization and the extinction ratio, we fit the following equation with the Levenberg-Marquardt algorithm, 
\begin{equation}
  I(\phi, \lambda) = I_{{\rm pol}}(\lambda) \sin2(\phi - \phi_{0}(\lambda)) + I_{{\rm mean}}(\lambda),
\label{eq:pol_def_lab}
\end{equation}
where $I_{{\rm pol}}$ is the brightness of the polarized component, 
$I_{{\rm mean}}$ indicates the mean brightness of the light source,
and $\phi_{0}(\lambda)$ represents the phase angle of polarization.
The degree of polarization, $P_{{\rm pol}}$ is calculated by the following equation,
\begin{equation}
P_{\rm pol} = \frac{I(\phi_{0}+45^\circ)-I(\phi_{0}-45^\circ)} {I(\phi_{0}+45^\circ)+I(\phi_{0}-45^\circ)} = \frac{I_{\rm pol}}{I_{\rm mean}}.
\label{eq:degree_of_polarization}
\end{equation}

We show the $P_{\rm pol}$ and $P_{\rm unpol}$ in Figure \ref{fig:lab_test}.
Although the $P_{\rm pol}$ for parallel light was larger than 0.999, and $P_{\rm unpol}$ is then less than 0.001,
$P_{\rm pol}$ measured in the experiment is $\approx$ 0.99. It is due to the oblique incident light at the position where the polarization film is installed in the LRS. However, $P_{\rm pol}$ is acceptably small to measure the polarization of ZL and we regard the unpolarized component of 0.01 offset as a systematic uncertainty in the LRS data.

\section{BASIC DATA REDUCTION}
\begin{figure}[bp]
\epsscale{1.2}
\plotone{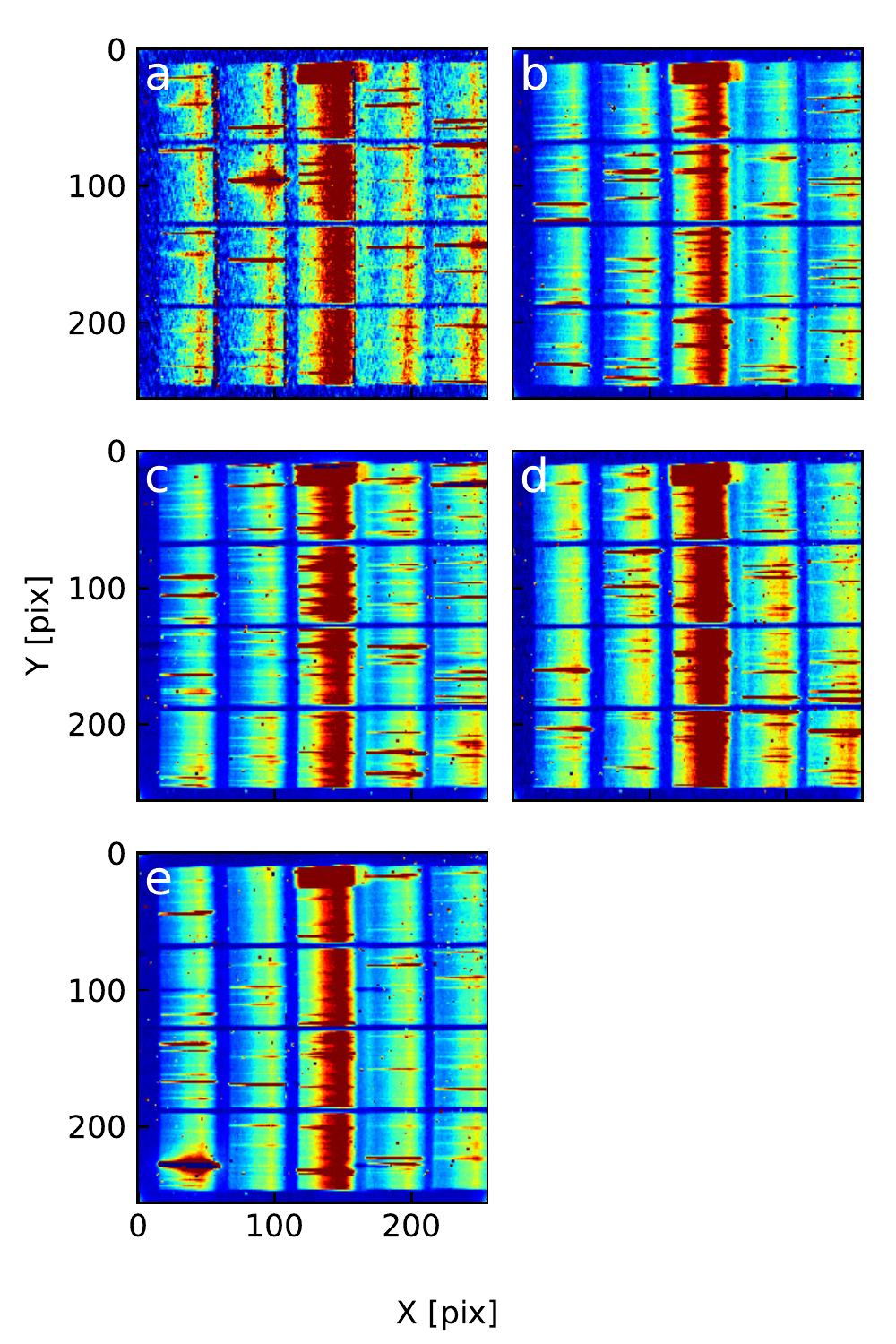}
\caption{LRS raw images of the CIBER third flight in the photocurrent unit taken in each observation field (a: Lockman, b: SWIRE/ELAIS-N1, c: NEP, d: Elat30, e: BOOTES-B). The five vertical sections correspond to the images of the five slits dispersed toward the $x$ direction, and four parallel sections correspond to wire-grid polarizing films.
}
\label{fig:sky_images}
\end{figure}
At first, we make the LRS image from individual time-ordered array readout frames.
In Figure \ref{fig:sky_images}, we give examples of the LRS images in the third flight.
The five vertical sections correspond to the five slits dispersed in the $x$ direction, and the four parallel sections correspond to the wire-grid polarizing films.

\begin{figure}[b]
\epsscale{1.2}
\plotone{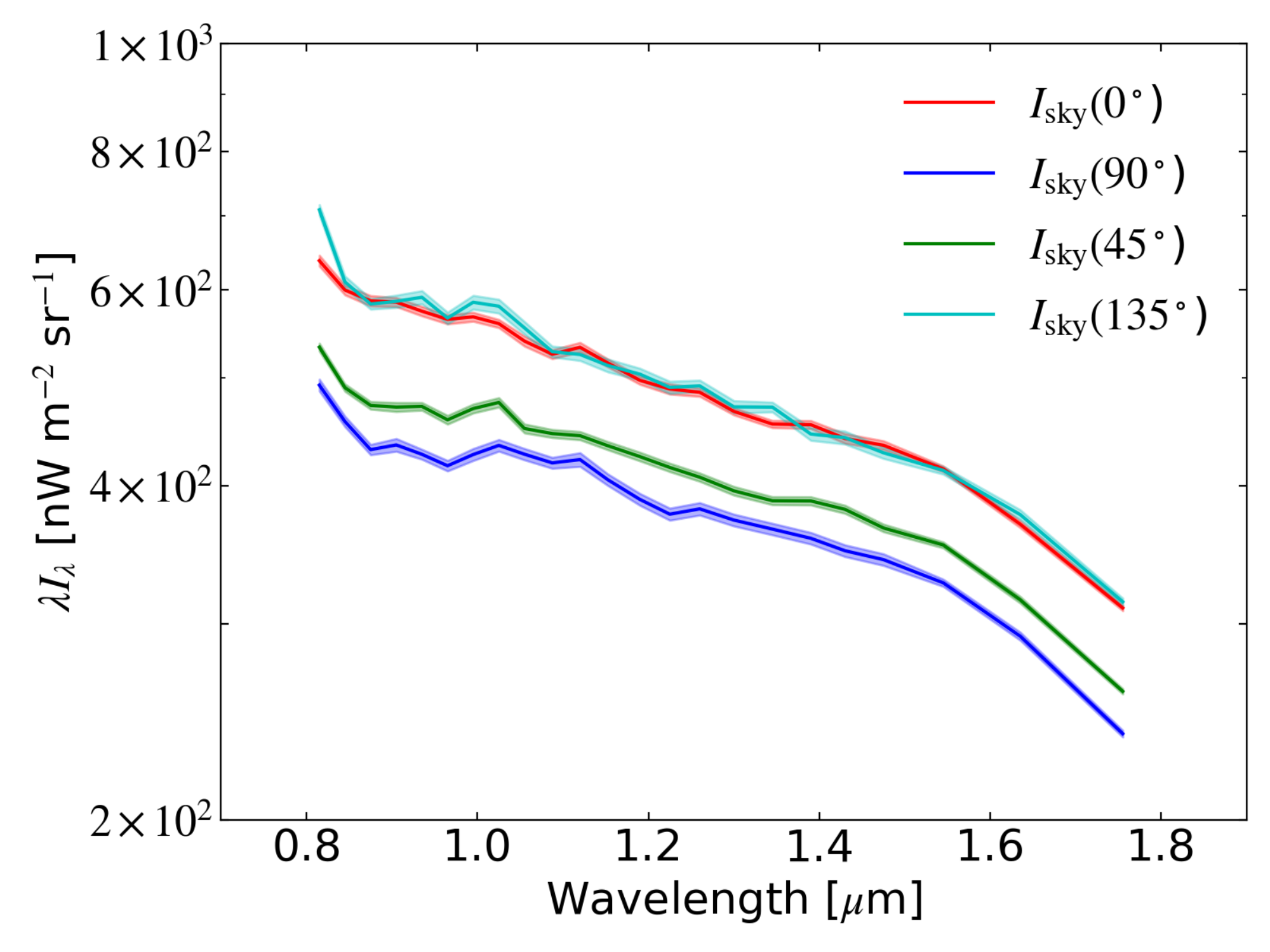}
\caption{The sky spectrum $I_{\rm sky}(\theta)$ through each polarization filter at the NEP field. The phase angles of the transmittance axis is labeled as $\theta$~=~$0^\circ, 45^\circ, 90^\circ, 135^\circ$.
The shaded regions dominated by the flux calibration uncertainty, and the statistical uncertainty is negligible.
}
\label{fig:nep_spec_paper}
\end{figure}

In order to measure the absolute spectrum of the sky brightness, we subtract the dark current and mask bright point sources from the LRS images.
The dark current is estimated from the masked regions of each slit individually.
The corners of the array are masked to avoid contamination by spurious signals emanating from the multiplexers of the detectors in the corners of each quadrant.
To remove bright point sources, we average the photocurrent of each slit along the horizontal direction, then clip the pixels containing stars determined by the criterion that the band-averaged photocurrent is deviated from the mean of band-average photocurrent of all pixels by 2 $\sigma$,
where $\sigma$ is the standard deviation of the photocurrent.
We iterate this clipping procedure until the ratio of the number of rejected pixels to the remaining pixels is less than 0.1$\%$ of the total.
To reject hot and dead pixels and the remaining faint point sources, pixels which are greater than 3 $\sigma$ from the mean are also excluded.
Finally, we derive the sky spectrum, $I_{\rm sky}(\theta)$, through each polarization filter as shown in Figure \ref{fig:nep_spec_paper}.

Further details on the basic data reduction are described in \citet{2010ApJ...719..394T} and \citet{arai2015measurements}.

\section{DATA ANALYSIS}

After the basic data reduction, to derive the ZL polarization spectrum, we separate the mean ZL brightness, $I_{\rm ZL, mean}$, and the ZL brightness of the polarized component, $I_{\rm ZL, pol}$, from the measured sky brightness. 
We assume that the polarization components of the DGL, the ISL, and the EBL are negligible compared to the ZL as discussed in Section \ref{seq:sys}.
The measured sky brightness of the polarized component $I_{{\rm sky, pol}}$ can then be assumed as $I_{{\rm sky, pol}}=I_{\rm ZL, pol}$.

The mean sky brightness comprises;
\begin{equation}
  I_{{\rm sky, mean}}(\lambda) = I_{{\rm ZL, mean}}(\lambda) +  I_{{\rm DGL}}(\lambda) + I_{{\rm ISL}}(\lambda) + I_{{\rm EBL}}(\lambda).
\label{eq:i_sky_mean}
\end{equation}
To derive $I_{\rm ZL, mean}$, the DGL, the ISL, and the EBL contributions need to be separated from the measured sky brightness \citep{matsuura2017new}.
The DGL component,  $I_{{\rm DGL}}(\lambda)$, is derived using its spatial distribution as traced by 100~$\mu$m emission on scales smaller than a degree \citep{arai2015measurements}.
The ISL component, $I_{{\rm ISL}}(\lambda)$, is estimated by Monte-Carlo simulation of the star distribution 
in the FOV using the 2MASS catalogue \citep{2006AJ....131.1163S}, 
taking into account the limiting magnitude and the effective slit efficiency of the LRS \citep{arai2015measurements}.
The EBL component, $I_{{\rm EBL}}(\lambda)$, can be assumed as identical in all the fields, but with a lot of indeterminacy.
To derive the fiducial unpolarimetric spectral shape of ZL, we calculate the difference between two fields, 
\begin{eqnarray}
    (I_{{\rm sky, mean}, i}(\lambda) -  I_{{\rm DGL}, i}(\lambda)&-&  I_{{\rm ISL}, i}(\lambda)) \nonumber  \\ - (I_{{\rm sky, mean}, j}(\lambda)  &-&  I_{{\rm DGL}, j}(\lambda)-  I_{{\rm ISL}, j}(\lambda)) \nonumber  \\
	 =(I_{{\rm ZL, mean}, i}(\lambda) +  I_{{\rm EBL}, i}(\lambda)) &-& (I_{{\rm ZL, mean}, j}(\lambda)  +  I_{{\rm EBL}, j}(\lambda)) \nonumber  \\
	 = I_{{\rm ZL, mean}, i}(\lambda) &-& I_{{\rm ZL, mean}, j}(\lambda)
\label{eq:ebl_cancel}
\end{eqnarray}
where $i$ and $j$ indicate different observation fields. 
We assume that the EBL component is identical in all the fields, so $I_{{\rm EBL}}(\lambda)$ is canceled in Equation \ref{eq:ebl_cancel}.
We also estimate the ZL brightness, $I_{\rm ZL_{model}}$, at 1.25 $\mu$m of each field from the DIRBE/COBE-based ZL model \citep{1998ApJ...508...44K}, which are summarized in Table \ref{tbl:field3}, and normalize the differences of $I_{\rm ZL, mean}$ with $I_{\rm ZL_{model}}$ at 1.25 $\mu$m.
Because of deep airglow contamination, only the NEP, Elat-30 and  BOOTES-B fields are available to derive the fiducial unpolarimetric spectral shape of ZL with polarization slit of the LRS.
To improve the signal-to-noise ratio, we use not only the data with polarization slit but also that with center slit at the NEP and Elat-30 fields, and average these results. 
It is noteworthy that we could not use the data with center slit at the BOOTES-B field because the detector was not functioning properly.
The airglow emission likely depends on atmospheric conditions and changes daily.
The details of uncertainty of the airglow contamination are described in Section \ref{seq:sys}.

\begin{figure}[tbp]
\epsscale{1.2}
\plotone{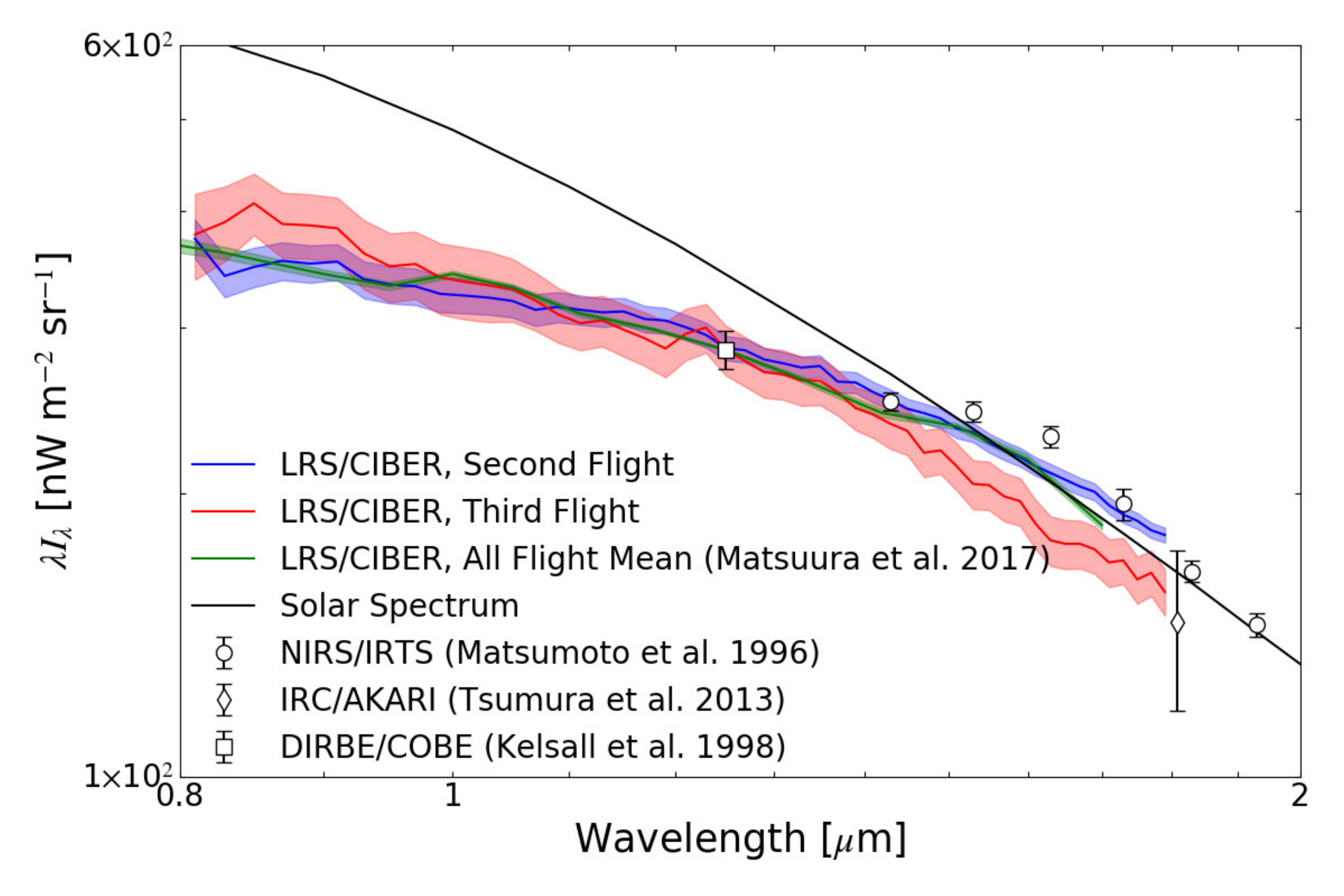}
\caption{The ZL spectrum measured in the second flight (blue line) and the third flight (red line). The solid green line presents the mean ZL spectrum of all CIBER flight \citep{matsuura2017new}. The shaded regions indicate 1$\sigma$ statistical uncertainties. The solid black line presents solar spectrum (http://rredc.nrel.gov/solar/spectra/am0). The circles indicate the ZL spectrum measured using NIRS/IRTS \citep{1996PASJ...48L..47M}. The diamonds present the ZL spectrum predicted by the IPD model derived from data from the Infrared Camera (IRC)/AKARI \citep{10.1093/pasj/65.6.121}. The square indicates the ZL brightness measured with DIRBE/COBE \citep{1998ApJ...508...44K}.
Since the target fields of the CIBER observations are different from those of other studies, they are scaled to the brightness estimated from the 1.25 $\mu$m ZL model \citep{1998ApJ...508...44K} at $(\lambda, \beta) = (335.37^\circ, 89.72^\circ)$.
}
\label{fig:ZL_spec}
\end{figure}

Figure \ref{fig:ZL_spec} presents the fiducial mean spectral shape of ZL during the third flight.
The difference in the ZL spectral shape between the second and third flights may be due to the timing of the observations and the calibration accuracy.
In deriving the ZL polarization spectrum, we only use the ZL spectral shape of the third flight in terms of the instrumental systematic uncertainties.

\begin{figure}[htbp]
\epsscale{1.2}
\plotone{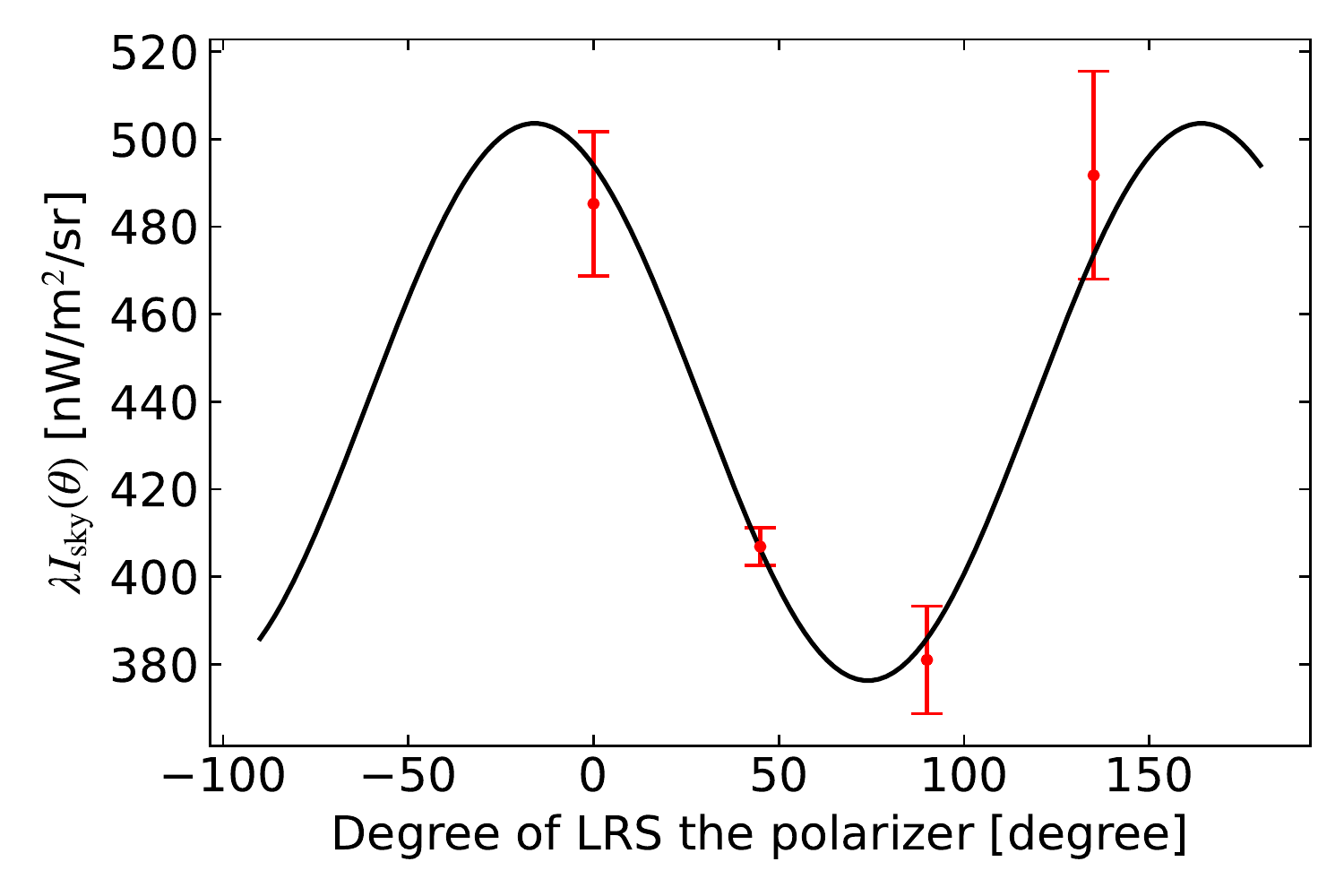}
\caption{The red circles indicate the measured sky brightness $I_{\rm sky}(\theta)$ at 1.26 $\mu$m through each polarization filter in the NEP field. The phase angles of the transmittance axis are labeled as $\theta$~=~$0^\circ, 45^\circ, 90^\circ, 135^\circ$.
The error bars are dominated by the flat-field error (see Section \ref{seq:sys1}).
The black line represents the best fit of Equation \ref{eq:pol_def_lab}.
$\lambda I_{\rm sky, mean}$ is $439\pm8$ nW m$^{-2}$ sr$^{-1}$ and $\lambda I_{\rm sky, pol}$ is $63\pm10$ nW m$^{-2}$ sr$^{-1}$.
}
\label{fig:nep_fit_1260}
\end{figure}

To derive $I_{\rm ZL, pol}$, we fit the measured sky spectrum $I_{\rm sky}$ \edit1{at all available wavelength} with Equation \ref{eq:pol_def_lab} to separate the polarized and unpolarized components. 
In Figure \ref{fig:nep_fit_1260}, we show the measured $I_{\rm sky}(\theta)$ at 1.26 $\mu$m as a function of $\theta$ in the NEP field, as an example of the analysis.
The black curve presents best fit by Equation \ref{eq:pol_def_lab}.

Finally, from Equation \ref{eq:degree_of_polarization}, the ZL polarization spectrum $P_{\rm ZL}(\lambda)$ can be expressed as, 
\begin{equation}
	 P_{\rm ZL}(\lambda) = \frac{I_{{\rm ZL, pol}}(\lambda)}{I_{\rm ZL, mean} (\lambda)}
\end{equation}

\section{RESULT}
\label{seq:result}
\subsection{Validity of the Measured Polarization Angle}
\begin{figure*}[t]
\epsscale{1.0}
\plotone{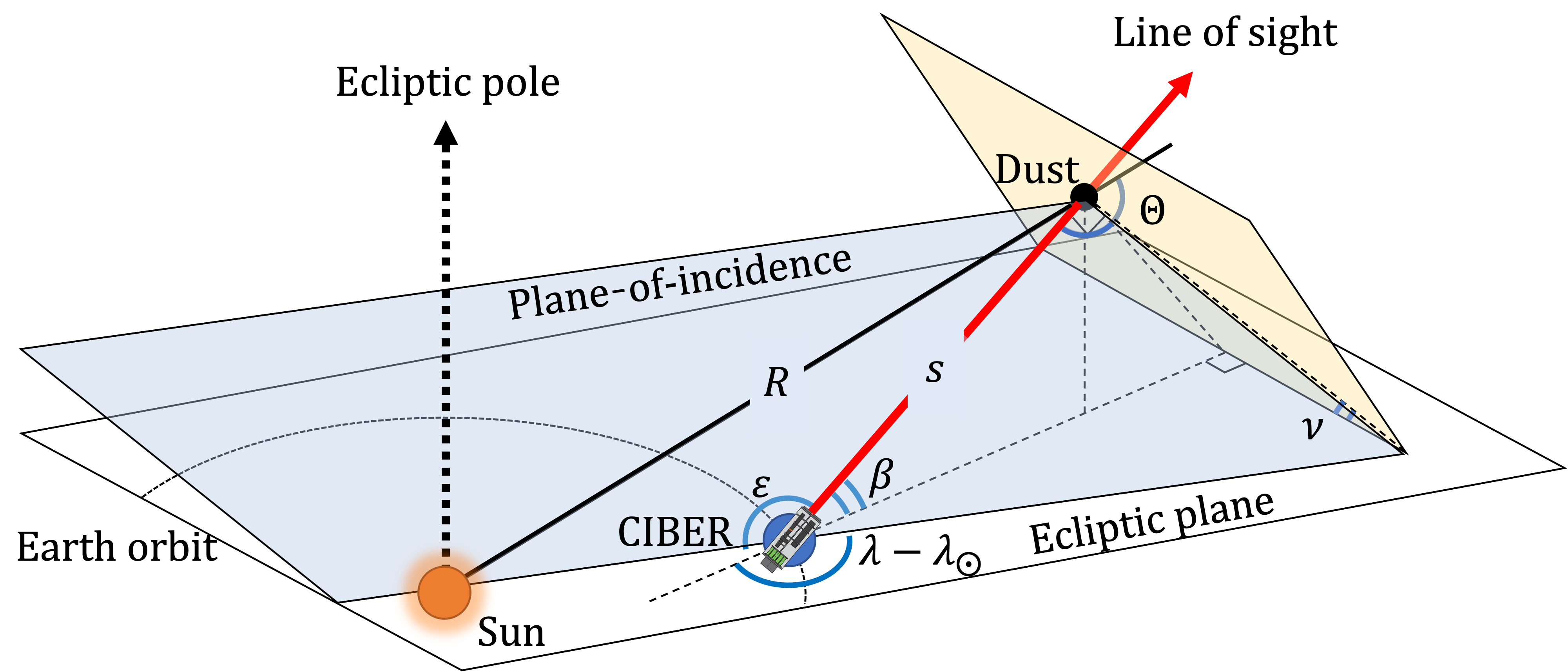}
\caption{The geometric description of CIBER observation with respect to the ecliptic latitude $\beta$ and the helio-ecliptic longitude ($\lambda - \lambda_\odot$), where $\lambda$ is the ecliptic longitude and $\lambda_\odot$ is  ecliptic longitude of the Sun.
The red arrow represents the line of sight from the LRS/CIBER, and $s$ indicates the distance between the LRS/CIBER and a IPD grain. 
\edit1{Observed ZL intensity corresponds to integrated light scattered by all IPD grains along the LRS/CIBER line of sight.}
The radial distance from the grain to the Sun is denoted by $R$, and the solar elongation is indicated by $\epsilon$.
The angle between the plane-of-incidence and the declination axis is represented by $\nu$.
The scattering angle is denoted by $\Theta$.
}
\label{fig:pol_geometry_1}
\end{figure*}

\begin{figure}[htbp]
\epsscale{1.0}
\plotone{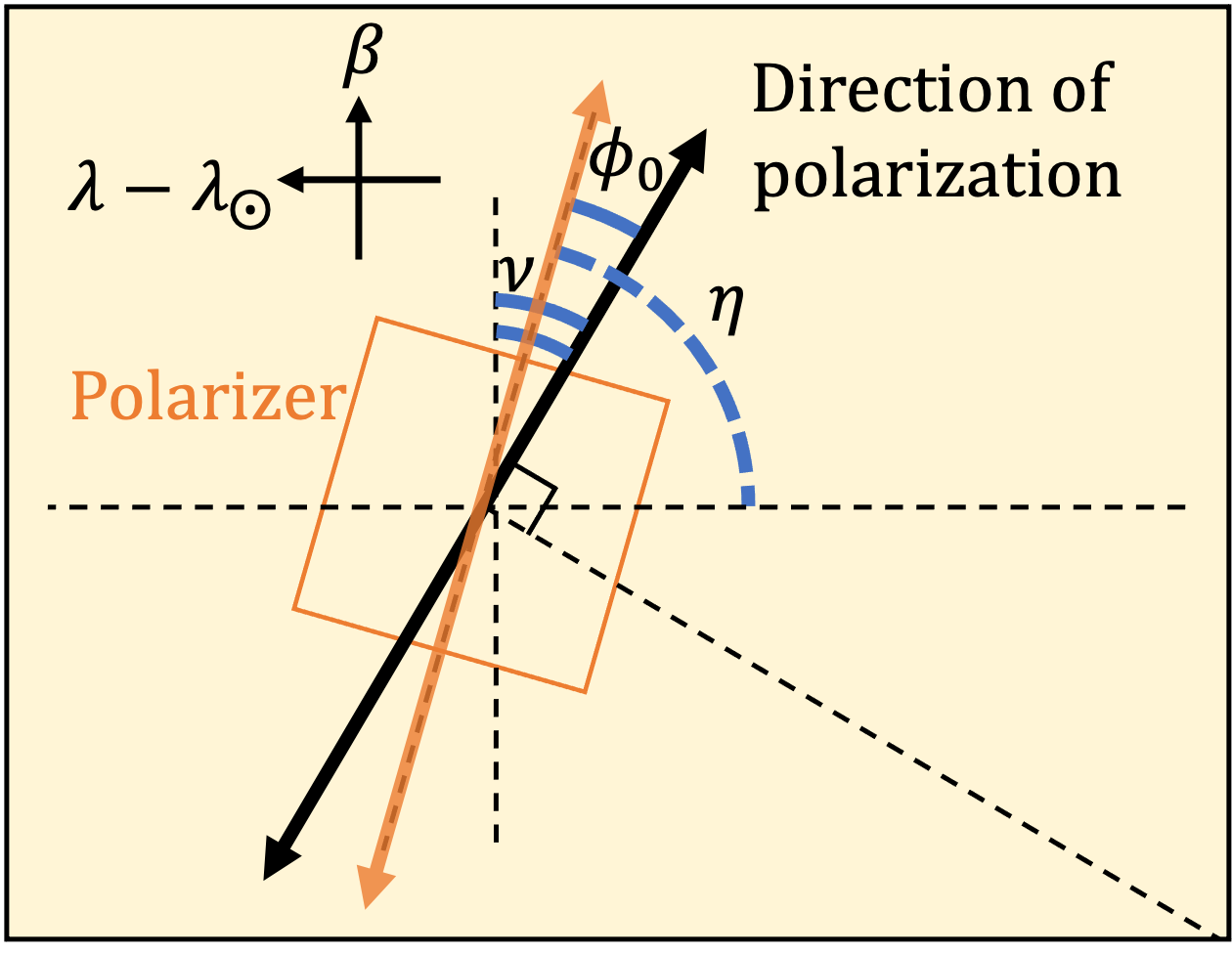}
\caption{The orientation of the LRS/CIBER polarizer relative to the direction of the ZL polarization, as viewed from the LRS toward the line of sight.
Helio-ecliptic coordinates are suitable to describe the ZL with the ecliptic latitude $\beta$ and the helio-ecliptic longitude ($\lambda - \lambda_\odot$).
The angle of plane-of-incidence toward the declination axis is represented by $\nu$.
The rotational angle of the polarizer toward the declination axis is indicated by $\eta$.
The angle of direction of polarization toward the transmission axis is labeled as $\phi_{0}$.}
\label{fig:pol_geometry}
\end{figure}

First of all, to validate our measurements, we compare an expected polarization angle, $\phi_{\rm 0,exp}$, with the measured polarization angle, $\phi_{\rm 0,meas}$.
We show the geometric definition of the polarization angle in Figure \ref{fig:pol_geometry_1} and \ref{fig:pol_geometry}. 
The angle of the plane-of-incidence, $\nu$, toward the declination is presented as
\begin{equation}
	 \nu  = \tan^{-1} \frac{ \sin(\beta)}{ \tan(180^\circ - \lambda + \lambda_\odot) }.
\end{equation}
Thus $\phi_{0}$ can be determined as
\begin{equation}
	 \phi_{0} = \eta + \nu - 90^\circ,
\end{equation}
where $\eta$ indicates the rotational angle of the slit. 
We note that the scattered light by each IPD grain along the line of sight produces the same phase angle since it does not depend on the distance between the LRS and the IPD grains.
We summarize these parameters to calculate the expected polarization angle in Table \ref{tbl:pol_angle}.
The measured phase angle is generally consistent with expectation, which supports our polarimetric measurements.

\begin{deluxetable}{ccccccc}
\tabletypesize{\scriptsize}
% \rotate
\tablecaption{The Expected Phase Angle $\phi_{\rm 0, exp}$ and the Measured Phase Angle $\phi_{\rm 0, meas}$}
\tablewidth{0pt}
\tablehead{\colhead{} &\phn&\phn&\colhead{$\phi_{\rm 0, exp}$} &\phn&\phn&\colhead{$\phi_{\rm 0, meas}$} \\
\colhead{Field Name} &\phn&\phn&\colhead{(deg)} &\phn&\phn&\colhead{(deg)}}
\startdata
Lockman Hole  &\phn&\phn& 113$\pm$1 &\phn&\phn& 118$\pm$12\\
SWIRE/ELAIS-N1 &\phn&\phn& 73$\pm$3 &\phn&\phn& 68$\pm$4\\
North Ecliptic Pole (NEP) &\phn&\phn& 120$\pm$1 &\phn&\phn& 119$\pm$2 \\
Elat30 ($\beta$ = 30 degree) &\phn&\phn& 49$\pm$1 &\phn&\phn& 54$\pm$8 \\
BOOTES-B &\phn&\phn& 190$\pm$2 &\phn&\phn& 163$\pm$8 \\
\enddata
\tablecomments{Because the LRS has large FOV, $\phi_{\rm 0, exp}$ is different between the center and the edge of an image.
The uncertainty of $\phi_{\rm 0, exp}$ indicates this difference.
$\phi_{\rm 0, meas}$ is the average over all wavelengths.
The uncertainty of $\phi_{\rm 0, meas}$ represents the standard deviation.
}
\label{tbl:pol_angle}
\end{deluxetable}

\subsection{Polarization Spectra}
\begin{figure}[htbp]
\epsscale{1.3}
\plotone{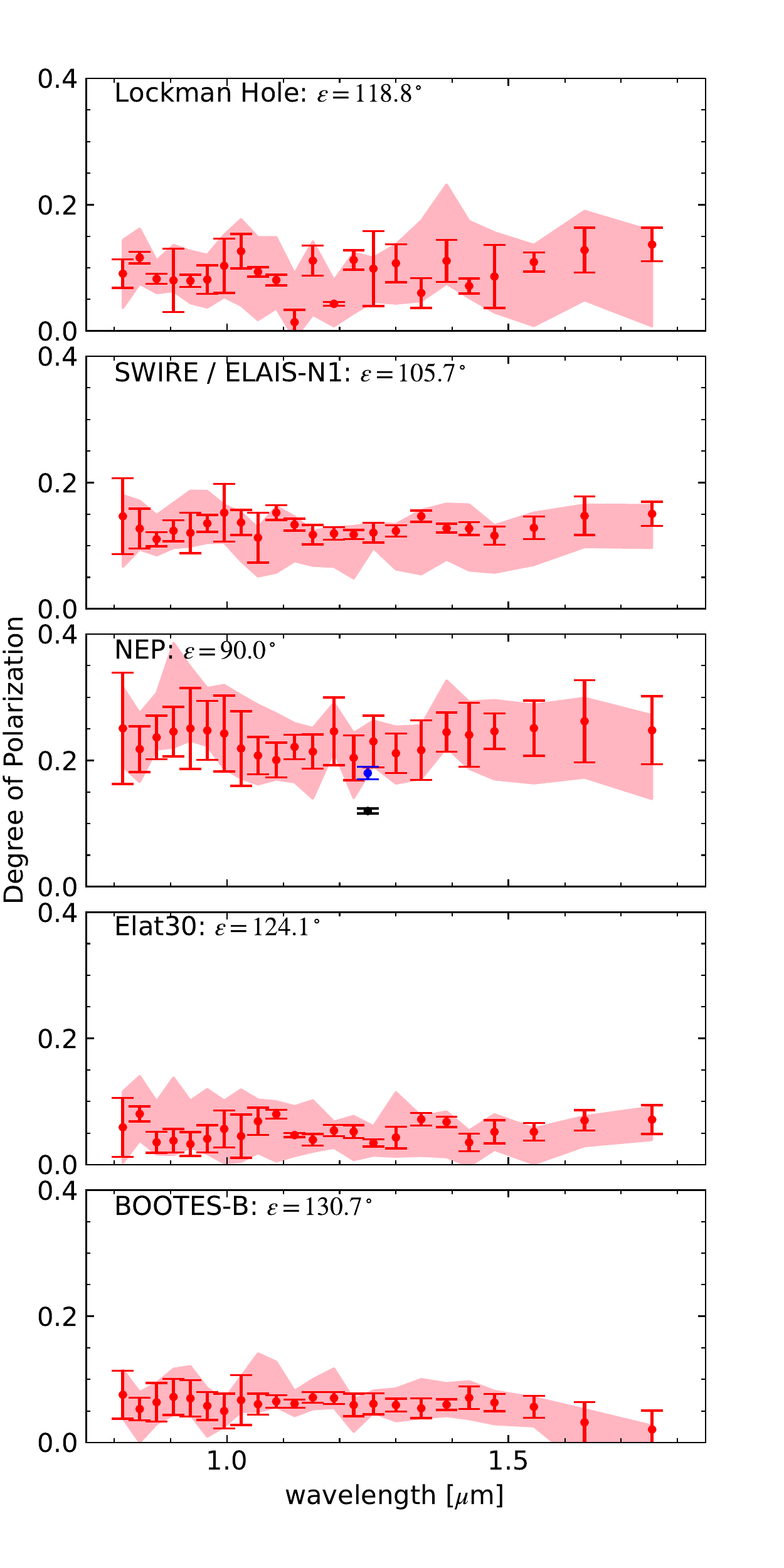}
\caption{The ZL polarization spectrum $P_{\rm ZL}(\lambda)$ measured in the third flight (red filled circles).
The error bars represent the total uncertainty due to the fitting of Equation \ref{eq:pol_def_lab} and the polarization calibration.
The red shaded region indicates the total instrument and astronomical systematic uncertainty (Section \ref{seq:sys}).
The black circle is the original data of the degree of polarization $P_{\rm sky, DIRBE}$ from \citet{1994ApJ...431L..63B}, and the blue circle is the corrected degree of polarization $P_{\rm ZL, DIRBE}$ measured with DIRBE/COBE at $\lambda = 10^\circ$ and $\beta = 0^\circ$.
}
\label{fig:pol_spec}
\end{figure}

We show the ZL polarization spectrum, $P_{\rm ZL}(\lambda)$, of the five fields measured in the third flight in Figure \ref{fig:pol_spec} and Table \ref{tbl:pol_spec}.
This is the first measurements of the polarization ZL spectrum in the near-infrared.
$P_{\rm ZL}(\lambda)$ shows little wavelength dependence in all fields.
On the other hand, $P_{\rm ZL}(\lambda)$ clearly depends on the ecliptic coordinates and the solar elongation.
$P_{\rm ZL}(\lambda)$ peaks at the NEP field where the solar elongation is $90^{\circ}$.
The detail of systematic uncertainty is described in Section \ref{seq:sys}.

\begin{deluxetable*}{cccccc}
\tabletypesize{\scriptsize}
% \rotate
\tablecaption{The ZL polarization spectrum $P_{\rm ZL}(\lambda)$ measured in the third flight.}
\tablewidth{0pt}
\tablehead{\colhead{$\lambda$}  
&\colhead{$P_{\rm ZL_{\rm Lockman~Hole}}$} &\colhead{$P_{\rm ZL_{\rm SWIRE/ELAIS\mathchar`-N1}}$}
&\colhead{$P_{\rm ZL_{\rm NEP}}$}
&\colhead{$P_{\rm ZL_{\rm Elat30}}$}
&\colhead{$P_{\rm ZL_{\rm BOOTES\mathchar`-B}}$}
\\
\colhead{($\mu$m)}
&\colhead{(\%)}
&\colhead{(\%)}
&\colhead{(\%)}
&\colhead{(\%)}
&\colhead{(\%)}
}
\startdata
0.815 &9.4 $\pm$ 2.2 $+$ 5.3/$-$5.5&15.2 $\pm$ 6.1 $+$ 3.4/$-$8.0&26.0 $\pm$ 8.8 $+$ 6.6/$-$6.0&6.1 $\pm$ 4.8 $+$ 3.4/$-$8.0&7.8 $\pm$ 3.9 $+$ 4.0/$-$3.7\\
0.845&13.1 $\pm$ 0.5 $+$ 4.6/$-$4.2&14.3 $\pm$ 3.4 $+$ 4.4/$-$3.3&24.5 $\pm$ 3.7 $+$ 5.7/$-$5.2&9.0 $\pm$ 1.2 $+$ 4.4/$-$3.3&6.0 $\pm$ 2.0 $+$ 2.7/$-$5.3\\
0.875&8.9 $\pm$ 0.6 $+$ 2.9/$-$2.3&11.9 $\pm$ 0.9 $+$ 3.8/$-$2.6&25.5 $\pm$ 3.3 $+$ 7.0/$-$1.9&3.8 $\pm$ 1.8 $+$ 3.8/$-$2.6&6.9 $\pm$ 3.2 $+$ 3.1/$-$3.5\\
0.905&8.6 $\pm$ 5.4 $+$ 5.6/$-$1.7&13.3 $\pm$ 1.6 $+$ 4.4/$-$2.7&26.3 $\pm$ 3.8 $+$ 14.1/$-$2.5&4.1 $\pm$ 2.0 $+$ 4.4/$-$2.7&7.7 $\pm$ 3.0 $+$ 4.5/$-$2.8\\
0.935&8.4 $\pm$ 0.8 $+$ 4.7/$-$3.5&12.7 $\pm$ 3.2 $+$ 6.7/$-$2.2&26.4 $\pm$ 6.4 $+$ 9.8/$-$1.9&3.4 $\pm$ 2.0 $+$ 6.7/$-$2.2&7.4 $\pm$ 3.0 $+$ 5.1/$-$2.6\\
0.965&8.5 $\pm$ 2.3 $+$ 3.9/$-$4.4&14.2 $\pm$ 0.9 $+$ 5.1/$-$3.1&25.9 $\pm$ 4.5 $+$ 6.7/$-$2.5&4.3 $\pm$ 2.2 $+$ 5.1/$-$3.1&6.1 $\pm$ 2.3 $+$ 2.8/$-$4.9\\
0.995&10.7 $\pm$ 4.4 $+$ 4.9/$-$5.0&15.8 $\pm$ 4.6 $+$ 1.3/$-$4.8&25.1 $\pm$ 5.9 $+$ 7.7/$-$5.6&5.9 $\pm$ 3.0 $+$ 1.3/$-$4.8&5.2 $\pm$ 2.8 $+$ 1.8/$-$2.5\\
1.025&13.0 $\pm$ 2.6 $+$ 5.1/$-$8.6&14.1 $\pm$ 1.7 $+$ 1.6/$-$6.1&22.5 $\pm$ 5.8 $+$ 8.5/$-$4.7&4.7 $\pm$ 3.5 $+$ 1.6/$-$6.1&6.9 $\pm$ 4.0 $+$ 3.7/$-$1.9\\
1.055&9.6 $\pm$ 0.4 $+$ 5.5/$-$7.6&11.6 $\pm$ 4.0 $+$ 1.6/$-$6.1&21.3 $\pm$ 2.7 $+$ 8.1/$-$4.6&7.0 $\pm$ 2.2 $+$ 1.6/$-$6.1&6.2 $\pm$ 1.7 $+$ 8.0/$-$1.2\\
1.087&8.0 $\pm$ 0.7 $+$ 6.8/$-$4.5&15.1 $\pm$ 0.7 $+$ 1.0/$-$9.5&20.0 $\pm$ 2.5 $+$ 7.4/$-$3.1&7.9 $\pm$ 0.5 $+$ 1.0/$-$9.5&6.5 $\pm$ 0.9 $+$ 6.3/$-$0.9\\
1.120&1.4 $\pm$ 1.9 $+$ 7.7/$-$2.6&13.1 $\pm$ 0.6 $+$ 1.3/$-$5.8&21.8 $\pm$ 1.5 $+$ 3.8/$-$5.6&4.6 $\pm$ 0.2 $+$ 1.3/$-$5.8&6.1 $\pm$ 0.5 $+$ 2.0/$-$2.0\\
1.152&10.8 $\pm$ 2.2 $+$ 3.1/$-$8.6&11.4 $\pm$ 1.3 $+$ 0.6/$-$4.9&20.7 $\pm$ 2.3 $+$ 3.8/$-$7.4&3.8 $\pm$ 0.9 $+$ 0.6/$-$4.9&6.9 $\pm$ 0.7 $+$ 2.9/$-$1.9\\
1.190&4.1 $\pm$ 0.1 $+$ 3.8/$-$3.6&11.3 $\pm$ 0.6 $+$ 1.0/$-$5.3&23.4 $\pm$ 4.9 $+$ 4.5/$-$3.2&5.1 $\pm$ 0.8 $+$ 1.0/$-$5.3&6.7 $\pm$ 0.8 $+$ 4.7/$-$1.6\\
1.225&11.4 $\pm$ 1.4 $+$ 1.1/$-$8.5&11.9 $\pm$ 0.4 $+$ 1.3/$-$7.0&20.7 $\pm$ 3.4 $+$ 3.9/$-$6.4&5.3 $\pm$ 1.0 $+$ 1.3/$-$7.0&6.0 $\pm$ 1.8 $+$ 0.9/$-$4.3\\
1.260&9.8 $\pm$ 5.9 $+$ 1.7/$-$5.3&12.0 $\pm$ 1.4 $+$ 1.5/$-$2.4&22.8 $\pm$ 3.9 $+$ 3.3/$-$3.9&3.4 $\pm$ 0.5 $+$ 1.5/$-$2.4&6.1 $\pm$ 1.6 $+$ 2.1/$-$1.4\\
1.300&10.5 $\pm$ 2.9 $+$ 3.0/$-$6.4&12.1 $\pm$ 0.6 $+$ 1.1/$-$6.1&20.7 $\pm$ 2.9 $+$ 4.2/$-$4.8&4.2 $\pm$ 1.7 $+$ 1.1/$-$6.1&5.8 $\pm$ 1.0 $+$ 2.6/$-$2.6\\
1.345&5.9 $\pm$ 2.3 $+$ 11.4/$-$1.3&14.3 $\pm$ 0.4 $+$ 1.0/$-$9.2&21.1 $\pm$ 4.4 $+$ 3.9/$-$4.6&7.0 $\pm$ 0.9 $+$ 1.0/$-$9.2&5.3 $\pm$ 1.5 $+$ 4.6/$-$1.6\\
1.390&10.6 $\pm$ 3.1 $+$ 12.1/$-$3.6&12.2 $\pm$ 0.3 $+$ 3.9/$-$5.0&23.4 $\pm$ 2.7 $+$ 8.2/$-$2.5&6.5 $\pm$ 0.7 $+$ 3.9/$-$5.0&5.7 $\pm$ 0.8 $+$ 3.4/$-$1.8\\
1.430&6.8 $\pm$ 1.1 $+$ 10.3/$-$1.9&12.1 $\pm$ 0.7 $+$ 3.8/$-$6.6&23.0 $\pm$ 4.7 $+$ 5.3/$-$5.4&3.4 $\pm$ 1.3 $+$ 3.8/$-$6.6&6.8 $\pm$ 1.7 $+$ 2.6/$-$3.4\\
1.475&8.0 $\pm$ 4.6 $+$ 7.0/$-$5.7&10.8 $\pm$ 1.2 $+$ 1.6/$-$5.9&22.9 $\pm$ 2.3 $+$ 4.9/$-$7.6&4.8 $\pm$ 1.7 $+$ 1.6/$-$5.9&5.9 $\pm$ 1.2 $+$ 1.9/$-$3.4\\
1.545&10.0 $\pm$ 1.3 $+$ 2.7/$-$10.1&11.8 $\pm$ 1.5 $+$ 2.4/$-$5.9&23.0 $\pm$ 3.8 $+$ 3.7/$-$8.8&4.8 $\pm$ 1.2 $+$ 2.4/$-$5.9&5.2 $\pm$ 1.6 $+$ 1.5/$-$3.1\\
1.635&11.2 $\pm$ 3.0 $+$ 6.2/$-$7.9&12.9 $\pm$ 2.6 $+$ 1.8/$-$4.9&22.9 $\pm$ 5.5 $+$ 3.7/$-$8.9&6.1 $\pm$ 1.3 $+$ 1.8/$-$4.9&2.8 $\pm$ 2.8 $+$ 2.1/$-$7.1\\
1.755&12.2 $\pm$ 2.3 $+$ 2.1/$-$13.0&13.4 $\pm$ 1.5 $+$ 1.4/$-$5.4&22.1 $\pm$ 4.7 $+$ 2.4/$-$10.9&6.4 $\pm$ 2.0 $+$ 1.4/$-$5.4&1.9 $\pm$ 2.7 $+$ 0.6/$-$10.3\\
\enddata
\tablecomments{Mean value $\pm$ Statistical uncertainty $+$ Systematic uncertainty (upper/lower).
}
\label{tbl:pol_spec}
\end{deluxetable*}
\begin{deluxetable*}{cccccccc}
\tabletypesize{\scriptsize}
% \rotate
\tablecaption{Corrected degree of polarization measured by DIRBE/COBE
at $\lambda = 10^\circ$ and $\beta = 0^\circ$.
}
\tablewidth{0pt}
\tablehead{
\colhead{Wavelength} &
\colhead{$P_{\rm sky, DIRBE}$} &
\colhead{Detection} &
\colhead{$I_{\rm ZL, mean}$} &
\colhead{$I_{\rm ISL}$} &
\colhead{$I_{\rm DGL}$} &
\colhead{$I_{\rm EBL}$} &
\colhead{$P_{\rm ZL, DIRBE}$} \\
\colhead{} &
\colhead{$ = \frac{I_{\rm ZL, pol}}{I_{\rm sky, mean, DIRBE}}$} &
\colhead{Limit} &
\colhead{} &
\colhead{} &
\colhead{} &
\colhead{} &
\colhead{$ = \frac{I_{\rm ZL, pol}}{I_{\rm ZL, mean}}$} \\
\colhead{($\mu$m)} &
\colhead{($\%$)} &
\colhead{(Jy)} &
\colhead{(nW~m$^{-2}$~sr$^{-1}$)} &
\colhead{(nW~m$^{-2}$~sr$^{-1}$)} &
\colhead{(nW~m$^{-2}$~sr$^{-1}$)} &
\colhead{(nW~m$^{-2}$~sr$^{-1}$)} &
\colhead{($\%$)} 
}
\startdata
1.25 & 12.00 $\pm$ 0.4 & 15 & 831 $\pm$ 41 & 318 $\pm$ 15 & 10 $\pm$ 1 & 54 $\pm$ 17 & 18 $\pm$ 1 \\
2.2 & 10.00 $\pm$ 0.5 & 15 & 302 $\pm$ 15 & 257 $\pm$ 13 & 5 $\pm$ 1 & 28 $\pm$ 7 & 19 $\pm$ 1 \\
\enddata
\label{tbl:cobe_pol}
\tablecomments{See Section \ref{seq:result} for detail.}
\end{deluxetable*}
\subsection{Solar Elongation Dependence}
\begin{figure*}[tbp]
\epsscale{1.2}
\plotone{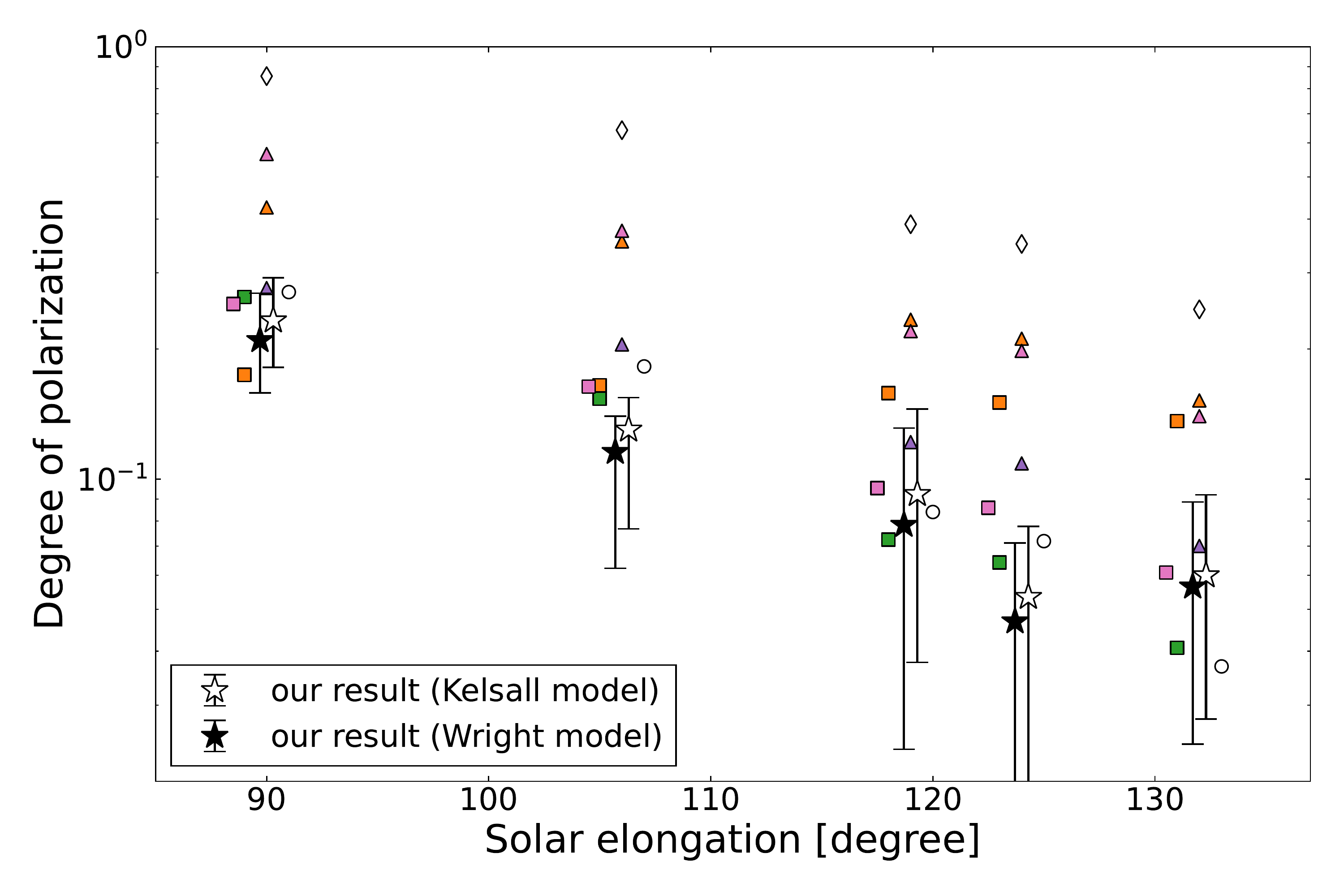}
\caption{
The degree of polarization $P_{\rm ZL}(\lambda)$ as a function of the solar elongation. 
Open and closed stars are average $P_{\rm ZL}(\lambda)$ between 0.8 and 1.8 $\mu$m scaled to the brightness estimated from the Kelsall model and the Wright model.
Error bars indicate the systematic uncertainties.
Open circles indicate the $P_{\rm ZL}$ produced by the empirical scattering in the visible band \citep{leinert1975zodiacal}, and open diamonds present that calculated by the Rayleigh scattering \citep{1983asls.book.....B} along the line of sight.
The triangles and the squares indicate the $P_{\rm ZL}$ calculated by the Mie scattering with astronomical silicate and graphite, respectively \citep{draine1984optical}.
The marker colors mean different particle radius (see Figure \ref{fig:dop_sa_comparison}).
Note that each marker has an offset in the X-axis so that they do not overlap with each other, even if they have the same solar elongation.
The Empirical model and the Mie model with graphite can clearly reproduce the measurements better than the other models.
}
\label{fig:dop_se_comparison}
\end{figure*}

\begin{figure}[tbp]
\epsscale{1.2}
\plotone{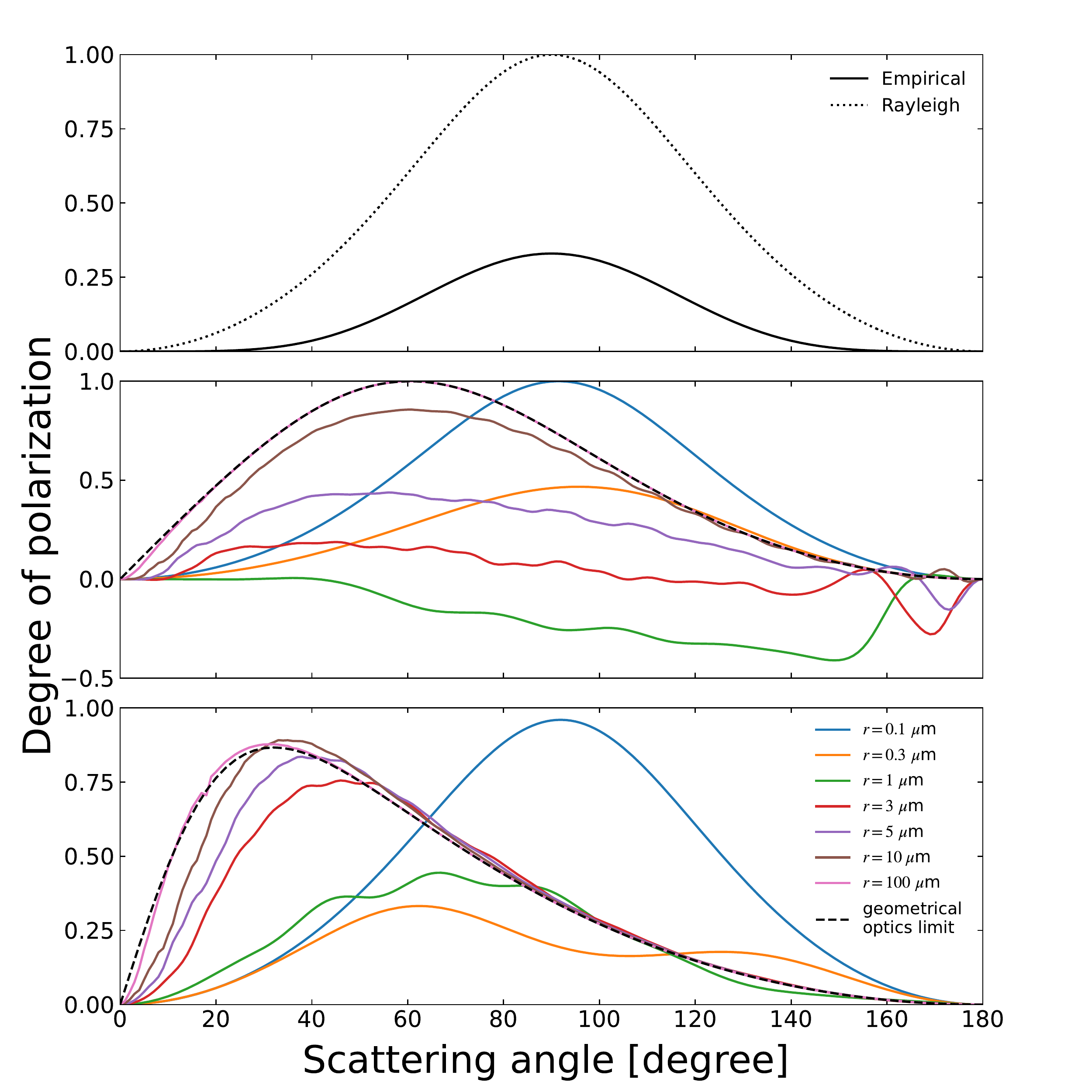}
\caption{
The degree of polarization as a function of the scattering angle $\Theta$. 
($Top$) The solid line indicates the empirical degree of polarization by the IPD in the visible band relative to $\Theta$ \citep{leinert1975zodiacal}.
The dotted line indicates the Rayleigh scattering.
($Middle$) The solid lines indicate the Mie scattering models using the complex refractive index of astronomical silicate.
The particle radius $r$ is varied from 0.1 to 100 $\mu$m, and each $r$ is integrated over $r\pm50\%$.
The dashed line represent the geometric optics limit.
($Bottom$) Same as middle, but using the complex refractive index of graphite.
}
\label{fig:dop_sa_comparison}
\end{figure}

We show the mean of $P_{\rm ZL}(\lambda)$ from 0.8 to 1.8 $\mu$m as a function of the solar elongation in Figure \ref{fig:dop_se_comparison}. 
We plot two results scaled to the ZL brightness estimated from the \citet{1998ApJ...508...44K} model (hereafter the Kelsall model) and the \citet{wright1998angular} model (hereafter the Wright model).
The degree of polarization of the \edit1{spherical} IPD as a function of the scattering angle $\Theta$ in different scattering models is shown in Figure \ref{fig:dop_sa_comparison}.
The models we consider are empirical scattering, Rayleigh scattering, and Mie scattering with astronomical silicate and with graphite \citep{1983asls.book.....B}.
\edit1{The size parameter $x$ is a dimensionless parameter that characterizes the particle's interaction with the incident radiation as
$$
    x=\frac{2\pi r}{\lambda},
$$
where $r$ is the particle's radius, $\lambda$ is the wavelength of the light.
The empirical scattering refers to the empirical degree of polarization of IPD in the visible band, $P=0.33 \sin^{5}\Theta$ \citep{leinert1975zodiacal}, but this function does not cover a sufficient range of particle sizes and materials to replace the Mie calculation.
The Rayleigh scattering applies to the case when the scattering particle is very small ($x \ll 1$, with $r<0.1\lambda$) and the whole surface re-radiates with the same phase.
At the intermediate $x \approx 1$ of Mie scattering, interference effects develop through phase variations over the object's surface.
An object with $x \gg 1$ scatters light according to its projected area, based on geometrical optics scattering.}
For Mie scattering, we assume the mean complex refractive indices $m=1.714 + 0.031i$ for astronomical silicate as typical refractive material and $m=2.995 + 1.590i$ for graphite as a typical absorptive material from 0.8 to 1.8 $\mu$m \citep{draine1984optical}.
\edit1{The middle and bottom of Figure \ref{fig:dop_sa_comparison} show Mie scattering models using the complex refractive index of astronomical silicate and graphite, respectively.
Small particles at $r=0.1~\mu$m ($x\gg1$) are approximated to Rayleigh scattering, while large particles at $r=100~\mu$m ($x\ll1$) are approximated to the geometric optics limit.}
The degree of polarization in each model is calculated by integrating along the line of sight in each field and assuming the IPD geometry of the Kelsall model (Figure \ref{fig:dop_se_comparison}).
Since each observation field in CIBER is toward solar elongation $\epsilon > 90^\circ$, the degree of polarization with scattering angle $< 90^\circ$ is not included in the line-of-sight integration.
Our results are consistent with the degree of polarization inferred from the scattering properties of the empirical model and the Mie scattering model with $r\geq1~\mu$m graphite.
The Rayleigh scattering and the Mie scattering with astronomical silicate can be rejected as the principal mechanism of the ZL polarization.
\edit1{If the order of wavelength and particle radius is comparable, the Mie scattering model shows complex scattering characteristics due to interference, and there are cases where the degree of polarization is close to our results.
The degree of polarization of the Mie scattering model with $r = 5~\mu$m atronomical silicate is similar to our results, but it does not make sense to explain the ZL polarization only by a certain dust size.}

\subsection{Comparison with the DIRBE/COBE data}
To compare our result with the DIRBE/COBE data at 1.25 and 2.2 $\mu$m,
we estimate the ZL polarization, $P_{\rm ZL, DIRBE}$, from the result presented in \citet{1994ApJ...431L..63B}.
Because \citet{1994ApJ...431L..63B} did not take into account the DGL, the ISL, and the EBL contributions,
they present the polarization of the sky brightness, $P_{\rm sky, DIRBE}$.
In order to estimate $P_{\rm ZL, DIRBE}$, we estimate $I_{\rm ZL, mean}$, $I_{\rm ISL}$, $I_{\rm DGL}$, and $I_{\rm EBL}$.
We estimate $I_{\rm sky, mean, DIRBE}$ as the following:
\begin{equation}
	I_{\rm sky, mean, DIRBE} = I_{\rm ZL, mean} + I_{\rm ISL} + I_{\rm DGL} + I_{\rm EBL},
\end{equation}
where $I_{\rm ZL, mean}$ is estimated from the Kelsall model.
$I_{\rm ISL}$ is calculated by integrating stars fainter than the detection limit of point sources of DIRBE, at 15 Jy \citep{1998ApJ...508...74A}.
We integrate star light fainter than 15 Jy by using 2MASS catalogue \citep{2006AJ....131.1163S} and TRILEGAL, which is population synthesis code for Monte-Carlo simulating a star count in the Galaxy \citep{2005A&A...436..895G}.
The uncertainty of $I_{\rm ISL}$ is estimated from the variance due to the Monte-Carlo simulation.
$I_{\rm DGL}$ is estimated from \citet{arai2015measurements} and \citet{10.1093/pasj/65.6.120} at 1.25 and 2.2 $\mu$m.
$I_{\rm EBL}$ is adopted from \citet{2001ApJ...555..563C}.
These estimated components are summarized in Table \ref{tbl:cobe_pol}.
Since these sources can be assumed as unpolarized, we calculate the surface brightness of the polarization component, $I_{\rm ZL, pol, DIRBE}$ as
\begin{equation}
	I_{\rm ZL, pol, DIRBE} = P_{\rm sky, DIRBE} I_{\rm sky, mean, DIRBE}.
\end{equation}
Finally we infer $P_{\rm ZL, DIRBE}$ using the estimated $I_{\rm ZL, mean}$ and $I_{\rm ZL, pol, DIRBE}$:
\begin{equation}
	P_{\rm ZL, DIRBE} = \frac{I_{\rm ZL, pol, DIRBE}}{I_{\rm ZL, mean}}.
\end{equation}
The resultant $P_{\rm ZL, DIRBE}$ is summarized in Table \ref{tbl:cobe_pol}.
Although the original DIRBE/COBE data, $P_{\rm sky, DIRBE}$, indicates redder color than our data,
$P_{\rm ZL, DIRBE}$ shows little wavelength dependence after we account for other diffuse sources.
We show comparison between $P_{\rm sky, DIRBE}$, $P_{\rm ZL, DIRBE}$, and $P_{\rm ZL}(\lambda)$ at the NEP field with the same solar elongation $\epsilon = 90^{\circ}$ in Figure \ref{fig:pol_spec}. $P_{\rm ZL, DIRBE}$ is consistent with $P_{\rm ZL}(\lambda)$ at the NEP field.
\section{SYSTEMATIC UNCERTAINTY}\label{seq:sys}

\begin{figure}[htbp]
\epsscale{1.3}
\plotone{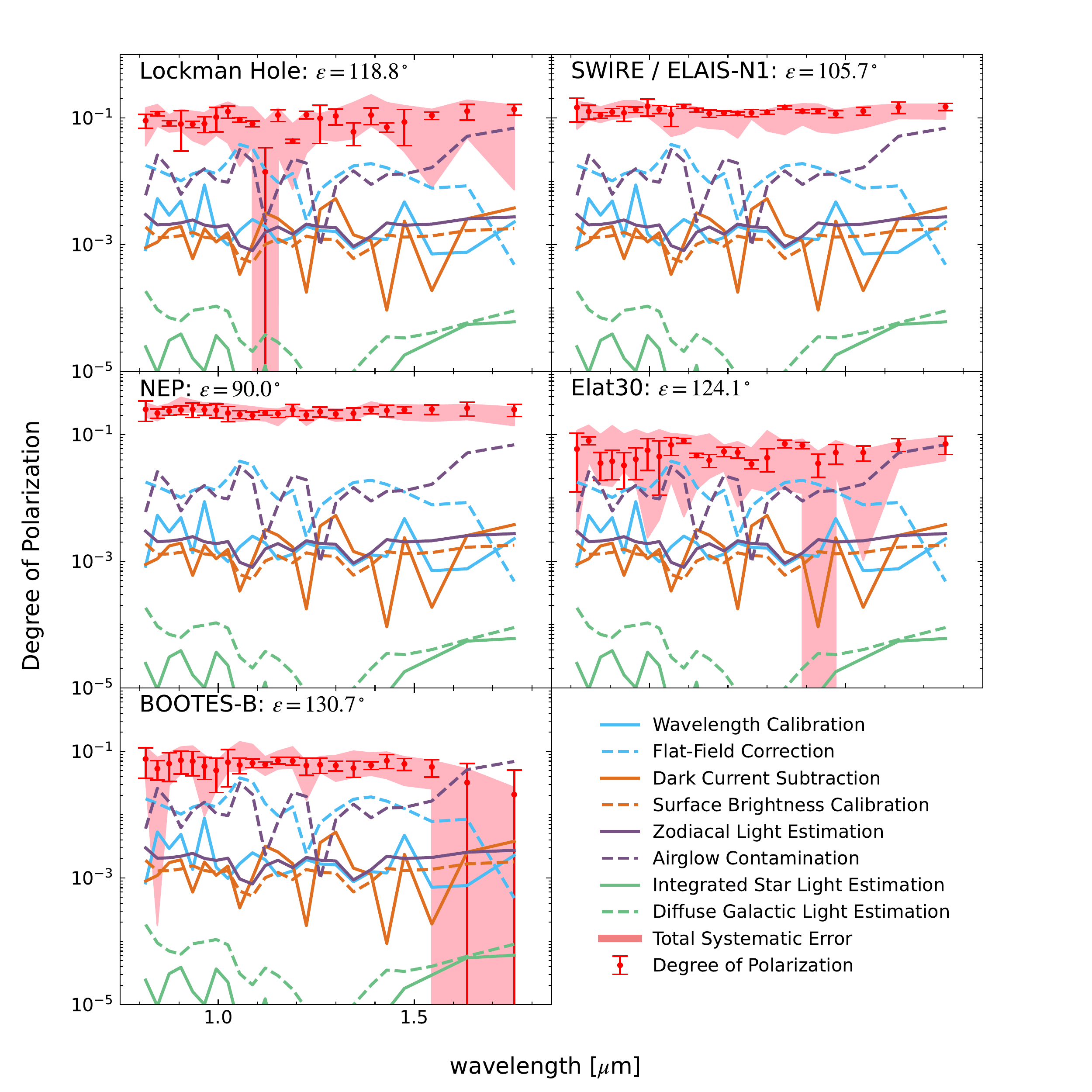}
\caption{The ZL polarization spectrum $P_{\rm ZL}(\lambda)$ measured in the third flight (red filled circles).
The instrumental systematic uncertainties associated with the wavelength calibration (aqua solid line), the flat-field (aqua dashed line), the dark current subtraction (orange solid line), 
and the surface brightness calibration (orange dashed line) are indicated.
The systematic uncertainties from the astronomical foreground composed by ZL (purple solid line), the airglow contamination (purple dashed line),
the ISL (light-green solid line), 
and the DGL (light-green dashed line) are also presented.
The red shaded region indicates the total instrument and astronomical systematic uncertainty that is the quadrature sum of the systematic uncertainties.
}
\label{fig:pol_spec_sys}
\end{figure}

Fig \ref{fig:pol_spec_sys} contains the same data as Fig \ref{fig:pol_spec}, but overplotted with the instrumental and astronomical systematic uncertainties.
The systematic uncertainties include the instrumental calibrations,
the contributions from airglow emission, and the diffuse brightness estimation.
The total systematic uncertainty is the upper limit because we assume maximum polarization of residual faint stars, and the DGL.

\subsection{Instrumental Systematic Uncertainty}\label{seq:sys1}
Since the accuracy of the wavelength calibration is $\pm$1 nm, to quantify how the wavelength uncertainty propagates into the ZL polarization spectrum, we shift the wavelength by $\pm$1 nm, then recompute the ZL polarization spectrum with the shifted wavelength in each case. The uncertainty due to the wavelength calibration is then captured by the difference between the $+1$ and $-1$ shifted ZL spectra.
As shown in Figure \ref{fig:pol_spec_sys}, the propagated uncertainty is negligible.

% \subsection{Flat-Field Correction}
We estimated the biased introduced by the flat-field correction by differencing the flat-field measurements from various integrating spheres.
We used two integrating spheres with 10~cm and 20~cm exit port diameters as described in \citet{arai2015measurements}.
We then derive two ZL spectra for these two flat fields, and take their difference as the uncertainty of ZL polarization due to the flat field. 
The flat-field systematic uncertainty is $<6$\%, which is acceptably small.
% \subsection{Dark Current Subtraction}
To check the systematic uncertainty from the dark current subtraction, we closed a cold shutter and measured the dark current images twice, on the rail and during the flight.
The flight-rail difference makes a $\sim$ 0.03 e- s$^{-1}$ systematic offset corresponding to $\sim$ 0.7 nW m$^{-2}$ sr$^{-1}$ at 1.25 $\mu$m.
We subtracted these dark current images from the each sky image and then calculated difference between the rail and flight cases, which gives us the bias from dark current subtraction.
The systematic uncertainty in the dark current subtraction does not significantly affect the final results.

% \subsection{Absolute Brightness and Polarization Calibration}
In general, the absolute brightness calibration uncertainty is canceled because the calibration is common between the polarization and the brightness channels.
However, since we estimate the ZL brightness using a ZL model, a bias arises when comparing our data with the ZL model brightness.
As described in \citet{arai2015measurements}, we conduct the absolute brightness calibration with several different setups in the laboratory.
The measured calibration factors are consistent to within a 3$\%$ r.m.s variation, which sets the uncertainty of the absolute brightness calibration.

To quantify the bias of the polarization measurement, we investigate the LRS polarization calibration data described in Section \ref{seq:cal}.
Although the extinction ratio of the wire-grid polarization film is higher than 1000, the measured extinction ratio of the LRS is 100.
The opening angle of incidence at the polarization film in the LRS is large, so some incident light leaks and decreases the extinction ratio of the LRS.
This means that the polarization measurement has a 0.01 offset.
We regard this offset as the systematic uncertainty of the polarization measurement by the LRS.

\subsection{Airglow Contamination}
We attribute a time and a rocket-altitude dependence to the airglow emission, so the observed brightness is written as
\begin{equation}
	 I_{obs}(t, h) = I_{sky} + I_{air}(t, h),
\end{equation}
where $I_{air}$ indicates the brightness of airglow emission, 
$t$ is the time from the launch, and $h$ is the altitude of the rocket.
To estimate the contamination from airglow, we separate the sky image into a first and a second-half integrations and derive the ZL polarization spectrum from each half.
If there is difference between the ZL polarization spectra derived from the first and the second-half images,
this difference should be due to the airglow contamination.
We calculate this difference as the bias from the airglow.
The measured bias from the airglow contamination is $\sim$~20$\%$ at the Lockman and Elat30 fields, $\sim$~10$\%$ at the BOOTES-B field, $\sim$~5$\%$ at the SWIRE field, and $\sim$~2$\%$ at the NEP field.
From these results, we confirm that the airglow contamination decays with time and altitude of the rocket, because the amount of the airglow contamination in the first-half is higher than that in the second half.
Because the integration time of Elat30 is shorter than other fields, the large airglow contamination can also be due to noise.
As shown in Figure \ref{fig:pol_spec_sys}, 
the airglow contamination is not negligible but is acceptably small.

\subsection{Astronomical Systematic Uncertainty}
Starlight is known to be linearly polarized due to interstellar dust grains aligned by the magnetic field of the Galaxy \citep{2000AJ....119..923H}.
The maximum polarization is 0.03 at $\tau_{V} =  1$ where $\tau_{V}$ is optical depth in V-band \citep{1973IAUS...52..145S}.
If the brightness of residual faint stars is $\sim$~15$\%$ of the ZL brightness at the NEP field, the maximum starlight polarization is 0.0045.
Because stars are randomly distributed and the LRS FOV is large, the starlight polarization should be lower than this estimation.
The starlight polarization is then only a few percent of the ZL polarization, which is negligible.

% \subsection{Diffuse Galactic Light}
Because the DGL polarization has never been measured, we adopt the polarization of molecular clouds and reflection nebulae to quantify the DGL polarization.
\citet{2007PASJ...59..481H} measured the linear polarization of the Orion molecular clouds and reported that the maximum polarization is 0.10 and 0.19 at J-band and H-band, respectively.
\citet{1987IAUS..115..374N} report that the polarization of the extended reflection nebulae around GGD27 IRS and W75N IRS is $\sim$~0.2 at 2.2~$\mu$m.
From these results, we assume that the DGL polarization is $\sim$~0.2.
Because the DGL brightness is $\sim$~10$\%$ of the ZL brightness at the NEP field, the contamination by the DGL polarization is 0.02.

\section{DISCUSSION}\label{seq:dis}
% \subsection{Size of Interplanetary Dust}
\begin{figure}[bp]
\epsscale{1.2}
\plotone{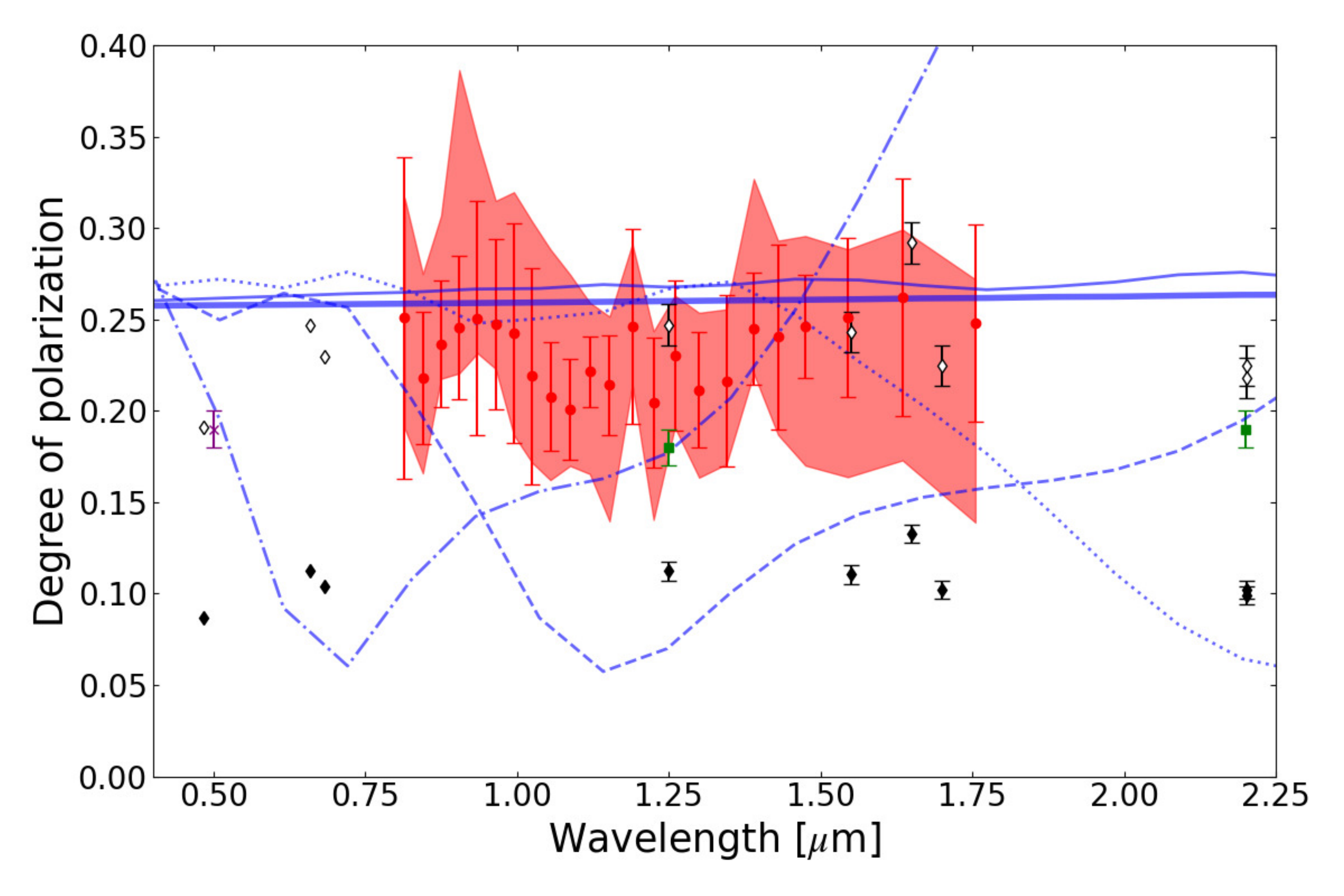}
\caption{Comparison between the ZL polarization spectrum, the ZL polarization of five Mie models, and the comet polarization.
The red filled circles indicate the ZL polarization spectrum $P_{\rm ZL}(\lambda)$ at the NEP field.
The error bars represent the total uncertainty from fitting of Equation \ref{eq:pol_def_lab} and the polarization calibration.
The red shaded region indicates the total systematic uncertainty.
\edit1{The green squares are the corrected degree of polarization $P_{\rm ZL, DIRBE}$ measured with DIRBE/COBE at $\lambda = 10^\circ$ and $\beta = 0^\circ$ (see Table \ref{tbl:cobe_pol}).}
The purple asterisk is the ZL polarization at 0.5~$\mu$m in the NEP field \citep{1980A&A....84..277L}.
The black diamonds are the polarization of comet Hale-Bopp extrapolated to 40$^\circ$ phase angle, the white diamonds are the same data, scaled by the purple asterisk \citep{2004AJ....127.2398K}.
The blue lines indicate the ZL polarization of five Mie scattering (spherical particle) models.
Thick solid line: $r=10~\mu$m. Thin solid line: $r=3~\mu$m. 
Dotted line: $r=1~\mu$m.
Dashed line: $r=0.5~\mu$m.
Dash-dot: $r=0.3~\mu$m.
}
\label{fig:zl_asteroid_comet}
\end{figure}
In Figure \ref{fig:zl_asteroid_comet}, the ZL polarization has little wavelength dependence from visible to infrared.
The similar scattering properties at different wavelengths suggest that large IPD grains contribute to the ZL polarization.
This suggestion is supported by the fact that our observations are consistant with the ZL polarization calculated by the empirical scattering model in the visible band.
The similar scattering properties at different wavelengths suggest that large IPD grains contribute to the ZL polarization.
We provide another test for the IPD size by comparing the ZL polarization spectrum with the polarization of Mie scattering theories.
We used five models with different particle radius ($x$=0.3, 0.5, 1, 3, and 10 $\mu$m) with a constant complex refractive index independent of wavelength, referring to $m=2.98+1.6i$ for graphite, a typical absorber.
The wavelength dependence of the ZL polarization from visible to infrared is consistent with the Mie scattering model for large particles ($r>1~\mu$m).
The observed degree of polarization is also roughly consistent with the Mie scattering model for large particles within the systematic uncertainty.

As shown in Figure \ref{fig:dop_se_comparison}, our results can also be reproduced by the Mie scattering with graphite, suggesting that the IPD is dominated by large particles ($r>1~\mu$m).
\edit1{However, graphite is not an significant component of the IPD complex and could be a better analogue for the very evolved organics detected in meteorites and comets \citep{HADAMCIK2020104527}.}
The observed ZL polarization is not consistent with the Rayleigh scattering which would favor small IPD grain size \edit1{($x\ll1$)}.
The Rayleigh scattering implies that the amount of scattering is inversely proportional to the fourth power of the wavelength.
However, the ZL polarization spectrum shows little wavelength dependence, therefore disfavoring the Rayleigh explanation.
In fact, the IPD is composed of several of these components, suggesting that the larger particles contribute more to ZL polarization than the smaller ones.
\edit1{We have used the spherical IPD model as a comparison, but more complex shapes such as aggregates and spheroids have been considered in recent ZL scattering models \citep{lasue2007inferring,kimura2016light}.
Our comparison results explain that IPD particles are not small spheres.}

% \subsection{Comparison with Cometary Dust}
In Figure \ref{fig:zl_asteroid_comet}, we also consider the polarization of \edit1{the cometary dust of} comet Hale Bopp extrapolated to 40$^\circ$ phase angle.
The polarization of cometary dust also has little wavelength dependence.
This is consistent with our observed polarization and that of large absorbing materials, however, cometary dust also contains silicate.
As reported by \citet{2009ASPC..418..395O}, the ZL spectrum exhibits crystalline silicate features at 9.3 and 11.35 $\mu$m.
Silicate emission bands similar to comets were also detected in the ZL spectrum \citep{2003Icar..164..384R}.
\citet{1999SSRv...90...99H} also reported Mg-rich silicate features at 9.3 and 11.3 $\mu$m.
\citet{zubko2014dust} reproduced the polarization measurement in comet C/1975 V1 (West) \edit1{in the visible band} by modeling a mixture of weakly and highly absorption particles with complex refractive indices $m=1.5 + 0i$ or $1.6 + 0.0005i$ and $2.43 + 0.59i$.
This mixture corresponds to Mg-rich silicates and amorphous carbon.
\edit1{Crystalline silicate features have been observed in the dust inner the coma of comet C/1995 O1 Hale-Bopp \citep{wooden1999silicate,wooden2000mg}. }
\citet{2009Icar..199..129L} also found evidences of a similar mixture comprising non-absorbing silicate-type and absorbing organic-type materials in the cometary dust of comet C/1995 O1 Hale-Bopp and 1P/Halley.

% \subsection{Future Work}
We cannot conclusively determine the origin of the IPD from our data alone.
\edit1{Future work is necessary to better constrain the properties of the IPD particles scattering the ZL.}
Further observation of comets and asteroids as well as theoretical modeling of the IPD scattering are required to reveal the origin of the IPD.
\edit1{Investigating the polarization properties of asteroidal dust is one of the most important future works.
In future model simulations, careful consideration of the variation of the complex refractive index of the astronomical material with wavelength will provide a better understanding of the structural and physical properties of the IPD \citep{dorschner1995steps}.
In addition, more detailed incorporation of complex particle shapes and size distributions into the model would allow us to reproduce the observed ZL polarization spectrum.}

We will extend the ZL observation to shorter wavelengths using the second CIBER experiment, CIBER-2 \citep{Takimoto2020}.
CIBER-2 was successfully launched in mid-2021 and its data will provide ZL spectrum at 0.5-2.5 $\mu$m. 
Future observations of ZL polarization spectra in the visible and near-infrared will help us understand the structural and physical properties of IPD.
At the same time, the heliocentric IPD distribution will be observed by ZL observation from outside the Earth's orbit (0.7-1.5 au) by the Hayabusa-2 extended mission \citep{HIRABAYASHI20211533}. 
These missions will allow us to better probe the IPD origin.

\section{Summary}
To determine the size and composition of the IPD, we observed the linear polarization spectrum of the ZL at the near-infrared wavelengths from 0.8 - 1.8 $\mu$m with spectro-polarimetric function of the LRS/CIBER instrument. 
We subtracted the contributions of the ISL, the DGL, and the EBL from the total sky brightness to derive the ZL polarization spectrum.
The ZL polarization spectrum shows little wavelength dependence, and the degree of polarization shows clear dependence on the ecliptic coordinates and the solar elongation.
The measured degree of polarization and its solar elongation dependence are reproduced by the empirical scattering model in the visible band and also by a scattering model for large absorptive particles, while small particles cannot reproduce ($r<1~\mu$m).
All of our results suggest that the IPD is dominated by large particles.
The wavelength dependence of the ZL polarization in wide wavelength range including the visible band is similar to that of comet Hale-Bopp, \edit1{although further data are needed to constrain the compositions.}
\edit1{Additional work is needed for a better constrain on the particles properties.}
\begin{acknowledgments}
This work was supported by NASA APRA research grants NNX07AI54G, NNG05WC18G, NNX07AG43G, NNX07AJ24G, and NNX10AE12G.
Initial support was provided by an award to J.B. from the Jet Propulsion Laboratory's Director's Research and Development Fund. 
CIBER was supported by KAKENHI (20.34, 18204018, 19540250, 21340047, 21111004, 2111717, 26800112, and 15H05744) from the Japan Society for the Promotion of Science (JSPS) and the Ministry of Education, Culture, Sports, Science and Technology (MEXT). 
Korean participation in CIBER was supported by the Pioneer Project from Korea Astronomy and Space science Institute (KASI).

We would like to acknowledge the dedicated efforts of the sounding rocket staff at the NASA Wallops Flight Facility and the White Sands Missile Range, and also thank Dr. Allan Smith, Dr. Keith Lykke, and Dr. Steven Brown (NIST) for the laboratory calibration of LRS.
P.K. and M.Z. acknowledge support from a NASA Postdoctoral Fellowship, T.A. acknowledges support from the JSPS Research Fellowship for Young Scientists, and A.C. acknowledges support from an NSF CAREER award AST-0645427 and NSF AST-1313319.
This publication makes use of data products from the \textit{2MASS}, which is a joint project of the University of Massachusetts and the Infrared Processing and Analysis Center/California Institute of Technology, funded by the National Aeronautics and Space Administration and the National Science Foundation.
\end{acknowledgments}

\bibliography{sample631}{}
\bibliographystyle{aasjournal}

\end{document}